\pdfoutput=1
\documentclass[12pt]{article}
\usepackage{times,graphicx,theorem,epsfig,lscape,amsmath,dcolumn,multirow,epic,float,enumerate,comment,bm}
\usepackage{hyperref}
\usepackage{natbib}
\usepackage{amsmath,amssymb}
\usepackage[normalem]{ulem}

\newcommand{\blinding}[2]{#1}   

\theoremstyle{plain} 

\theoremstyle{plain}

\theoremstyle{plain} 
\theoremstyle{plain}

\theoremstyle{plain}

\newcommand{\one}{\mathbf{1}}

\DeclareMathOperator\bE{\mathbb E} 
\DeclareMathOperator\bV{\mathbb V} 

\newcommand{\bX}{\bm{X}}

\newcommand{\ATT}{\mbox{\tiny{ATT}}}

\newcommand{\ATO}{\mbox{\tiny{ATO}}}

\newcommand{\bx}{\bm{x}}
\newcommand{\bbeta}{\bm{\beta}}

\newtheorem{assumption}{Assumption}

\begin{document}

\begin{center}
\vspace*{-2cm}

{\Large A Regression Discontinuity Design for Ordinal Running Variables: Evaluating Central Bank Purchases of Corporate Bonds}

\medskip
\blinding{
Fan Li \quad Andrea Mercatanti \quad Taneli M\"akinen \quad Andrea Silvestrini
\footnote{FL is associate professor, Department of Statistical Science, Duke University, Durham, NC, 27708 (email: fl35@duke.edu); AM is researcher (andrea.mercatanti@liser.lu), Bank of Italy and Luxembourg Institute of Socio-Economic Research; TM (email: taneli.makinen@bancaditalia.it) and AS (email: andrea.silvestrini@bancaditalia.it) are researchers, Bank of Italy. The authors are grateful to Federico Apicella, Johannes Breckenfelder, Federico Cingano, Riccardo De Bonis, Alfonso Flores-Lagunes, Frank Li, Fabrizia Mealli, Santiago Pereda Fern\'andez, Stefano Rossi and Stefano Siviero for helpful comments and suggestions. Part of this work was done while TM was visiting the Einaudi Institute for Economics and Finance, whose hospitality is gratefully acknowledged. The views expressed herein are those of the authors and not necessarily those of Bank of Italy. All remaining errors are ours.}

\today
}
\date{}

\end{center}

\date{}

{\centerline{ABSTRACT}

\noindent Regression discontinuity (RD) is a widely used quasi-experimental design for causal inference. In the standard RD, the assignment to treatment is determined by a continuous pretreatment variable (i.e., running variable) falling above or below a pre-fixed threshold. In the case of the corporate sector purchase programme (CSPP) of the European Central Bank, which involves large-scale purchases of securities issued by corporations in the euro area, such a threshold can be defined in terms of an ordinal running variable. This feature poses challenges to RD estimation due to the lack of a meaningful measure of distance. To evaluate such program, this paper proposes an RD approach for ordinal running variables under the local randomization framework. The proposal first estimates an ordered probit model for the ordinal running variable. The estimated probability of being assigned to treatment is then adopted as a latent continuous running variable and used to identify a covariate-balanced subsample around the threshold. Assuming local unconfoundedness of the treatment in the subsample, an estimate of the effect of the program is obtained by employing a weighted estimator of the average treatment effect. Two weighting estimators---overlap weights and ATT weights---as well as their augmented versions are considered. We apply the method to  evaluate the causal effect of the CSPP 
and find a statistically significant and negative effect on corporate bond spreads at issuance. 


\clearpage


\section{Introduction}


Regression discontinuity (RD) is a widely used quasi-experimental design for causal inference. In the original sharp RD design, the treatment status is a deterministic step function of a pretreatment variable, commonly referred to as the \textit{running variable}. All units with a realized value of the running variable on one side of a pre-fixed threshold are assigned to one regime and all units on the other side are assigned to the other regime. The basic idea of RD is that one can compare units with similar values of the running variable, but different levels of treatment, to draw causal inference of the treatment at or around the threshold. First introduced in 1960 \citep{ThistlethwaiteCampbell1960}, RD has become increasingly popular since the late 1990s in economics and policy, and more recently in medical research, with many influential applications \citep[e.g.,][among others]{AngristKrueger1991, ImbensvanderKlaauw1995, AngristLavy1999, vanderKlaauw2002, mealli2012evaluating, geneletti2015bayesian}.

In the standard RD setting, the running variable is continuous; one usually assumes continuity (namely, potential outcomes are continuous functions of the running variable at the threshold), and then employs local linear regressions or polynomials to extrapolate the counterfactual potential outcome under the opposite treatment status and estimate the causal effects \emph{at} the threshold \citep[e.g.][]{HahnEtAl2001, ImbensLemieux2008, skovron2015practical, gelman2019high}. Recently, an increasingly popular strand of research instead contends that RD designs lead to locally randomized experiments \emph{around} the threshold \citep{Lee2008, LeeLemieux2010, CattaneoEtAl2015}. Building on this interpretation, several recent works provide formal identification conditions and inferential strategies \citep[e.g.,][]{CattaneoEtAl2015, LiEtAl2015, calonico2019regression}. {Despite the conceptual and operating difference between the two RD frameworks, a recent empirical comparison show their results on the same application are similar \citep{MatteiMealli2016}.}

In this article, we consider a setting where the running variable is ordered categorical (also referred to as ordinal). {This is a common setting in RD. For example, in 
financial markets as in our motivating application, purchases of bonds are often decided based on a cutoff of the rating of the bonds, which is an ordinal variable. In criminology, the inmate classification system is usually based on an ordered categorical security custody score (from least severe to most severe, calculated from the type of offenses) and an inmate's security level is assigned according to a set of cutoff points of this score \citep{berk1999evaluation, hjalmarsson2009juvenile}. In environmental studies, the United States Environmental Protection Agency uses an ordinal rating system to determine a site's level of danger to public welfare and order a cleanup based on a cutoff of the rating \citep{greenstone2008does}. In education, eligibility of programs is often based on letter-grade threshold. In surveys such as customer satisfaction surveys, a corrective intervention may be triggered if average response falls below a certain threshold in the ordinal Likert scale.}

A categorical, ordered or non-ordered, running variable poses challenges for RD estimation for two reasons. First, RD estimation usually involves measuring the distance of each unit to the threshold. When the running variable is categorical, the values of the running variable provide little information on the distance to the threshold. Consequently, one can no longer compare outcomes within arbitrarily small neighborhoods of the threshold to identify the causal effects, and thus has to account for the uncertainty about the relationship between the running variable and the outcomes \citep{LeeCard2008}. Second, if the number of categories is small, even considering only units in the two categories bordering the threshold may lead to misleading results, particularly when the units within the two categories differ considerably from each other.


It is important to differentiate the above categorical running variables from discretized ones, whose underlying variable is continuous, but is measured only over a set of moderate number of distinct values due to rounding. For example, the eligibility rule in many policies is based on age, which is often recorded only in years or quarters instead of days. The key difference is that discretized continuous variables have a meaningful scale of distance, whereas categorical variables do not. The existing RD literature for discrete running variables is limited and has mostly focused on discretized continuous variables. \citet{LeeCard2008} assume a parametric functional form relating the outcome to the running variable and account for the uncertainty in the choice of this functional form, and adopt cluster-consistent standard errors, but \citet{KolesarRothe2018} show that these confidence intervals can have poor coverage properties. \citet{Dong2015} shows that rounding continuous running variable in standard RD estimation leads to biased causal estimates and provides formulas to correct for this discretization bias. \citet{ImbensWager2018} develop an optimization-based approach that is applicable to both continuous and discrete running variables. Both \citet{KolesarRothe2018} and \citet{ImbensWager2018} proceed under the continuity perspective and assume that the finite set of observed values of the running variable is a subsample of the real line, and therefore a finite metric space. This allows one to interpolate the continuous outcome functions to the discrete observed values of the running variable. 
However, these methods are not directly applicable to cases with categorical running variables because the absence of a concept of distance in the topology of categorical variable excludes it from any class of finite metric spaces. 

In this paper, motivated by the evaluation of the European Central Bank's Corporate Sector Purchase Programme (CSPP) (Section \ref{sec:CSPP}), we develop a new method for conducting RD inference with ordinal running variables.  {We proceed under the local randomization framework to RD because it separates design (namely, finding a small neighborhood around the threshold) from analysis (namely, estimating the effect within the neighborhood), and thus is more flexible to accommodate complex situations such as ordinal running variables.}
Our proposal is a three-step procedure. First, we postulate a parametric (e.g. ordered probit) model for the ordinal running variable, and take the estimated probability of being assigned to a category above the threshold as the surrogate continuous running variable. Second, based on the estimated probability, we identify a subset of units in which the covariates in the treatment and control groups are similar. Third, within such a subset, we use the estimated probability to construct a weighted sample to estimate the causal effect of the treatment in a subpopulation \emph{around} the threshold. The weighted sample represents a population of interest, namely units that could conceivably have been assigned to either treatment status. 
This strategy is similar to the popular propensity score weighting in observational studies \citep{Hahn1998, Hirano2001, Hirano2003, LiEtAl2018}. 
{To improve the robustness and efficiency of the estimation, we further adopt the outcome-regression-augmented weighting estimators.} We also derive an M-estimator for the variance of the causal effect that incorporates the uncertainty arising from both the design and analysis stages. {Though motivated from the specific CSPP evaluation, the proposed method is readily applicable to all RD studies with ordinal running variables, including the aforementioned examples.  }

The rest of the paper proceeds as follows. Section \ref{sec:CSPP} describes the main institutional features of the CSPP.  Section \ref{sec:Methods} introduces the general framework, the methods and the estimation strategy. Section \ref{sec:Empirical application} describes the empirical application. Section \ref{sec:Conclusions} concludes.

\section{Motivating Application: The Corporate Sector Purchase Program} \label{sec:CSPP}

On March 10, 2016, the European Central Bank (ECB) announced the corporate sector purchase programme (CSPP), a new asset purchase program, to be implemented in conjunction with the other non-standard monetary policy measures already in place. Under the CSPP, the Eurosystem purchases investment-grade bonds issued by euro-area non-bank corporations. As a part of the ECB's expanded asset purchase programme (APP), the CSPP aims at strengthening the pass-through of the Eurosystem asset purchases to the financing conditions of the real economy, in pursuit of the ECB's price stability objective. Between its announcement and December 2018, the CSPP had a total cost of approximately 180 billion euros. Accurately evaluating the effects of the CSPP is crucial for future calibrations of the ECB's unconventional monetary policies.

For this application, our goal is to assess how the eligibility for purchase (i.e. treatment) under the CSPP affects bond spreads (i.e. outcome) at the time of their issuance. Here the treatment is a bond being eligible of purchase by the Eurosystem, and the assignment to treatment is determined by an ordinal variable -- the rating of that bond: only bonds with an investment-grade rating (i.e., BBB- or above) were eligible for purchase by the Eurosystem. {A conventional approach would be directly comparing the spreads of all investment-grade versus non-investment-grade bonds, adjusting for the pre-treatment covariates via regression or matching. The key underlying causal assumption is unconfoundedness, that is, the treatment is randomly assigned among all bonds conditional on the covariates. But this assumption is unlikely to hold given that the vast difference in the bonds across rating categories may not all be captured by the covariates. A local version of the unconfoundedness appears to be more plausible; thus, an alternative approach is RD with the bond rating as the running variable, which capitalizes on the locality of the treatment assignment. Specifically, one could compare the bonds in the category that is right above versus right below the threshold (BBB- versus BB+). However, such an approach can be jeopardized by the substantial heterogeneity within each rating category, as is the common problem with categorical running variables.}

{Credit agencies determine the rating of a bond by the financial strength of its issuer and bond-specific characteristics, based on some algorithms that usually involves cutoffs \citep[e.g.][]{Hickman1958, PogueSoldofsky1969, BlumeEtAl1998}. This observation motivates us to quantify the distance of each unit to the threshold in terms of a continuous latent (running) variable which determines the assignment of each unit to a category, and use the three-step RD procedure described earlier to evaluate the CSPP. In a sense, we circumvent the challenges of an ordinal running variable by reconstructing the rating process of a bond.}


There are a few earlier works analyzing the CSPP. \citet{Zaghini2019} assesses the effects of the program in the context of the primary bond market, by controlling for many possible determinants of bond spreads. Looking instead at the secondary market, \citet{AbidiFlores2018} employ differences in credit rating standards between investors and the ECB to shed light on the market's reaction to the announcement of the program. \citet{ArceEtAl2017} and \citet{Grosse-RueschkampEtAl2018}, conversely, investigate how the program affected bank lending. Finally, \citet{GalemaLugo2017} examine the individual bond purchases under the CSPP and their effects on the financing decisions of the issuers. We complement these works by providing estimates of the effect of the program which rely on a formal statistical framework of causal inference. Our application is also relatively novel as the RD framework has only recently been applied in the field of financial economics (examples include \citealp{Rauh2006}, \citealp{ChavaRoberts2008}, \citealp{KeysEtAl2010} and \citealp{BechtEtAl2016}).

\section{Methods}\label{sec:Methods}

\subsection{Setup and assumptions}
Consider a sample of $N$ units indexed by $i=1, \dots, N$ drawn from a super-population. Let $R_{i}:\{r_{1},..., r_{j}, ...,r_{J}\}$ be the ordinal running variable with $J$ categories and $r_{j} > r_{j-l}$ for any integer $1 \leq l \leq (j-1)$. Based on $R_i$, a binary treatment $Z_i$ is assigned according to an RD rule: if a unit has a value of $R_i$ falling above (or below, depending on the specific application) a pre-specified threshold, $r_t$, then that unit is assigned to treatment; otherwise, that unit is assigned to control. That is, the treatment status is given by $Z_i=\one(R_i \geq r_t)$, where $\one(\cdot)$ is the indicator function. For each unit, besides the running variable, a set of pretreatment covariates $\bX_i$ is also observed. Each unit has a potential outcome $Y_{i}(z_i)$ corresponding to each treatment level $z_i=0,1$, and only the one corresponding to the observed treatment status $Y_{i}=Y_{i}(z_i)$ is observed. Define the propensity score $e(\bx_{i})$ as the probability of unit $i$ receiving the treatment conditional on the covariates: $e(\bx_{i})\equiv \Pr(Z_{i}=1|\bX_{i}=\bx_i)=\Pr(R_{i} \geq r_{t}|\bX_{i}=\bx_i)$.

For valid causal inference, we focus on the subpopulations whose units all have non-zero probability of being assigned to either treatment condition. Formally, we make the assumption of \emph{local overlap}.

\begin{assumption}[Local overlap]\label{as::overlap} There exists a subpopulation $\Omega_{o} \subset \Omega$ such that, for each $i$ in $\Omega_{o}$, we have $0<e(\bX_i)<1$.
\end{assumption}

We will elaborate on the selection of this subpopulation $\Omega_{o}$ in Section \ref{sec:Select}. Within $\Omega_{o}$, we further make the following two assumptions.

\begin{assumption}[Local SUTVA]\label{as::SUTVA} For each unit $i$ in $\Omega_{o}$, consider two realizations of the running variable $r'_{i}$ and $r''_{i}$ with possibly $r'_{i} \neq r''_{i}$. If $z'_{i}=z''_{i}$, that is, if either $r^{'}_i \leq r_t$ and $r^{''}_i \leq r_t$, or $r^{'}_i > r_t$ and $r^{''}_i > r_t$, then $Y_{i}(z'_{i})=Y_{i}(z''_{i})$, irrespective of the realized value of the running variable $r_j$ for any other unit $j \neq i$ in $\Omega_{o}$.
\end{assumption}

Local SUTVA implies (i) the absence of interference between units, and (ii) the independence of the potential outcome on the running variable given the treatment status for the same unit.

\begin{assumption}[Local unconfoundedness]\label{as::unconfound}
For each unit $i$ in $\Omega_{o}$, the treatment assignment is unconfounded given $\bX_{i}$: $\Pr(Z_i|Y_{i}(1),Y_{i}(0), \bX_{i})=\Pr(Z_i|\bX_{i}).$
\end{assumption}
Local unconfoundedness underpins the randomization perspective of RD: it entails the existence of a subpopulation around the threshold for which the assignment to treatment is unconfounded given the observed pretreatment variables.  Local unconfoundedness is a weaker version of the \emph{local randomization} assumption in \citet{LeeCard2008}, and is similar to the \emph{bounded conditional independence} assumption in \citet{AngristRokkanen2012}. As explained by \citet[p.~655]{LeeCard2008}, the local randomization assumption means that ``it may be plausible to think that treatment status is `as good as randomly assigned' among the subsample of observations that fall just above and just below the threshold.'' {Local randomization is stronger than the standard continuity assumption as it concerns a small neighbor \emph{around} (instead of \emph{at}) the threshold; in fact, it implies continuity at the threshold.}
The local unconfoundedness assumption relaxes the local randomization assumption by allowing the probability to be assigned to the treatment to depend on the pretreatment variables. {This relaxation is important for two reasons: (i) local unconfoundedness is more plausible than local randomization in most real applications; and (ii) it allows us to enlarge the subsample of units around the threshold for which randomization can be assumed to hold. The latter also gives more flexibility in handling covariates. Note that our method in Section \ref{sec:proposal} still applies in the cases of no covariates, where local unconfoundedness simply reduces to local randomization. But covariates generally strengthen an RD analysis for the above reasons; furthermore, specific to our setting, covariates improve the goodness-of-fit of the parametric model of the forcing variable and reduce the standard errors of the estimates. Also see discussions in \cite{calonico2019regression}.} Our unconfoundedness assumption is ``local'' in nature: indeed, we maintain the RD hypothesis according to which, for each unit in the population, the treatment assignment also depends on the unobserved units' characteristics. As a result, the interval around the threshold for which the unconfoundedness hypothesis holds has to be bounded.

\subsection{Design and analysis} \label{sec:proposal}
The key to our proposal is to treat the probability $\Pr(R_{i}=r_{j}|\bX_{i})$ as a latent continuous running variable instead of using the observed category $R_i$ as an ordinal running variable. Under this perspective, we will define the causal estimand as a weighted average treatment effect on a subpopulation with particular policy interest, namely, the \emph{overlap population}.

Our estimation strategy consists of three steps. First, we fit a parametric model for the ordinal running variable conditional on the observed covariates and take the estimated probability of being assigned to treatment as the latent continuous running variable. Second, based on the estimated probability, we identify a subset of units in which the local unconfoundedness assumption is plausible by checking the covariate balance. Third, within this subpopulation, we estimate the average treatment effect for the target population.


%

\subsubsection{Probit model for the ordered running variable} \label{sec:Probit}
We postulate an ordered probit model for the distribution of the ordered running variable, $\Pr(R_{i}=r_{j}|\bX_{i}=\bx_i)$, and consequently for the propensity score $e(\bX_{i})$. Specifically, we assume that each unit's observed category $R_i$ is determined by a latent normally distributed variable $R_{i}^*$ as follows:
\begin{eqnarray}\label{eq::probit}
R_{i}^*=\bX_{i} \bbeta + \mathcal{E}_{i}, ~~~ \mathcal{E}_{i} \sim N(0,1)
\end{eqnarray}
and
\begin{equation} \label{eq::probit_cut}
R_i=\left\{
\begin{array}{ll}
r_1,   &\mbox{if } R_{i}^* \leq u_{1}, \\
r_{j}, &\mbox{if } u_{j-1} < R_{i}^* \leq u_{j},\\
r_{J}, &\mbox{if }  R_{i}^* > u_{J-1},\\
\end{array}
\right.
\end{equation}
where $u_{j} \in \{u_{0}, u_{1}, \ldots ,u_{J-1}, u_{J} \}$ is a series of cutoff points, with $u_{0} = -\infty$ and $u_{J} = \infty$. That is, $R_{i}$ falls in category $r_{j}$ when the latent variable $R_{i}^*$ falls in the interval between $u_{j-1}$ and $u_{j}$. A probit model for $R_{i}$ is plausible in contexts where the category classifies units by ordered levels of ``quality'', which can be for example the grade a student achieves in a subject, or in our case the credit quality of a bond.
In these examples, the quality of a unit is supposed to be a continuous variable (e.g., the student's level of knowledge in a subject, or the issuer's capacity to honor its debts) we cannot observe, but for which we can observe the interval where it falls. {
Indeed, the ordered probit model is common in the bond rating literature \citep{KaplanUrwitz1979, KaoWu1990, BlumeEtAl1998}.}  Based on the ordered probit model \eqref{eq::probit}--\eqref{eq::probit_cut}, we have
\begin{equation*}
\Pr(R_{i}=r_{j}|\bX_{i}=\bx_i)=\Pr(u_{j-1} < R_{i}^* \leq u_{j})=\Phi(u_{j} - \bx_{i} \bbeta) - \Phi(u_{j-1} - \bx_{i} \bbeta).
\end{equation*}

The ordered probit model belongs to the class of generalized linear models suitable for ordinal responses. The link function is the inverse of the normal CDF, which implies that the probability of response is a monotonic function of the linear transformation $\bx_{i} \bbeta$ \citep{AgrestiBook}, namely, for any $\bx_1$ and $\bx_2$, $\Pr(R_{i} \leq r_{j}|\bX_{i}=\bx_{1}) \leq \Pr(R_{i} \leq r_{j}|\bX_{i}=\bx_{2})$ when $\bx_{1} \bbeta > \bx_{2} \bbeta$. Given the deterministic relationship $Z_{i}=\one(R_{i} \geq r_{t})$, the monotonicity also holds for the propensity score $e(\bx_{i})$. Therefore, we expect the estimated propensity scores, $\hat{e}(\bx_{i})$, to be close to 1 for units for which we observe high values of $R_{i}$, while being close to 0 for units for which we observe low values of $R_{i}$.

Moreover, given the monotonicity of $e(\bx_{i})$ in $\bx_{i} \bbeta$, and provided that $\bX_{i}$ is a good predictor of the ordinal responses, we expect the average $\hat{e}(\bx_{i})$ to be below 0.5 for units whose value of $R_{i}$ is just below the threshold $r_{t}$, i.e., $\sum_{i} \one(R_i = r_{t-1}) \hat{e}(\bX_{i}) / \sum_{i} \one(R_i = r_{t-1}) <0.5$, and above or equal to 0.5 for units whose value of $R_{i}$ is at the threshold $r_{t}$, i.e., $\sum_{i} \one(R_i = r_{t}) \hat{e}(\bX_{i}) / \sum_{i} \one(R_i = r_{t}) \geq 0.5$. Therefore, values of the propensity score around 0.5 pertain to units which fall in categories around the threshold. These units form a target population of policy interest because they can be assigned with non-negligible probability to either treatment condition and therefore are the mostly affected by, even small, changes in the policy. This target population can be formally defined using the concept of ``overlap weights'', as described in Section \ref{sec:Estimands}.

In practice, a well-specified ordered probit model would produce in-sample predictions of $e(\bx_{i})$ that satisfy the above patterns, which can be verified by inspecting the box plots of the estimated $\Pr(R_i=r_j|\bX_{i}=\bx_{i})$ in each category of the observed running variable.


\subsubsection{Causal estimands and estimators} \label{sec:Estimands}
Within a subpopulation $\Omega_0$ where Assumptions \ref{as::overlap}--\ref{as::unconfound} hold, we can define a class of causal estimands over different target populations. Denote the density of the pretreatment variables in $\Omega_0$ by $f(\bx_{i})$, and the density of a target population by $g(\bx_{i})$, which is not necessarily the same as $f(\bx_{i})$. We call $h(\cdot)=g(\bx_{i})/f(\bx_{i})$ a tilting function. Also denote the conditional expectation of the potential outcome in treatment $z$ ($z=0,1$) in $\Omega_0$ by $\mu_z(\bx_{i})=\mathbb{E}\{Y_i(z)|\bX_i=\bx_{i}\}$. Then we can define the average treatment effect in the target population $g$ by a weighted average treatment effect (WATE) estimand \citep{Hirano2003}:
\begin{equation}
\tau_h=\mathbb{E}_g[Y_i(1)-Y_i(0)]=\frac{\mathbb{E}\{h(\bx_{i})(\mu_1(\bx_{i})-\mu_0(\bx_{i}))\}}{\mathbb{E}\{h(\bx_{i})\}}.
\end{equation}

For any given $h(\bx_{i})$, we can use the balancing weights---$w_{1}(\bx_{i})= h(\bx_{i})/e(\bx_{i})$ for the treated units and $w_{0}(\bx_{i})=h(\bx_{i})/(1-e(\bx_{i}))$ for the controls---to balance the distribution of the pretreatment variables between the groups \citep{LiEtAl2018}, and estimate the causal effect $\tau_h$. Specifically, a consistent moment estimator of $\tau_{h}$ is the sample difference in the weighted average outcomes between treatment groups
\begin{gather} \label{eq:estimator}
\hat{\tau}_{h}= \frac{\sum_{i} w_{1}(\bX_{i})Z_{i}Y_{i}}{\sum_{i} w_{1}(\bX_{i})Z_{i}} - \frac{\sum_{i} w_{0}(\bX_{i})(1-Z_{i})Y_{i}}{\sum_{i} w_{0}(\bX_{i})(1-Z_{i})}.
\end{gather}

We consider two specific target populations (and equivalently estimands and weights). The first target population is the overlap population, which has the most overlap in the pretreatment characteristics between the treatment and control groups. It is obtained by the overlap weights \citep{LiEtAl2018}: $\{(w_0=e(\bx_i), w_1=1-e(\bx_i))\}$, corresponding to $h(\bx_i)=e(\bx_i)(1-e(\bx_i))$, the maximum of which is attained at $e(\bx_i)=0.5$. The corresponding causal estimand is the \emph{average treatment effect for the overlap population (ATO)}. Arguably, the overlap population consists of the units whose treatment assignment might be most responsive to a policy shift as new information is obtained. In our RD framework, the \textit{overlap population} is exactly the subpopulation around the threshold: with overlap weights, the units are smoothly downweighted as their latent running variable moves away from the threshold, i.e., $e(\bx_i)=0.5$. The second target population is the treated population and the corresponding estimand is the average treatment effect for the treated (ATT). It is obtained by the ATT weights $\{w_0=e(\bx_i)/(1-e(\bx_i)), w_1=1\}$, corresponding to $h(\bx_i)=e(\bx_i)$, and is of common interest in the economic literature. 



{The moment weighting estimator \eqref{eq:estimator} can be sensitive to the specification of the propensity score model. This is particularly relevant in our application because considerable amount of subjectivity on bond ratings is well known in the literature  \citep{BlumeEtAl1998}. To improve robustness of the weighting estimators, we consider their augmented (by outcome regression model) versions. Specifically, let $\hat{\mu}_z(\bX_i)$ be the predicted value of $\mu_z(\bX_i)$ from a regression model for unit $i$. For the ATT, we adopt the augmented estimator proposed by \cite{mercatanti2014debit}:
\begin{equation} \label{eq:ATT_aug}
\hat{\tau}_{aug}^{\ATT}=\frac{\sum_{i} Y_i Z_i}{\sum_{i} Z_i} - \frac{1}{\sum_{i} Z_i}\sum_{i}\frac{Y_i(1-Z_i)\hat{e}(\bX_i)+\hat{\mu}_0(\bX_i)(Z_i-\hat{e}(\bX_i))}{1-\hat{e}(\bX_i)}.
\end{equation}
This estimator is doubly robust in the sense that it is consistent if either the propensity score model or the potential outcome model is correctly specified, but not necessarily both \citep{scharfstein1999adjusting}. For the ATO, we develop a new augmented estimator as follows:
\begin{eqnarray}\label{eq:ATO_aug}
\hat{\tau}_{aug}^{\ATO}&=&
\frac{\sum_{i}(1-\hat{e}(\bX_i)) \left(Z_i Y_i-(Z_i-\hat{e}(\bX_i))\hat{\mu}_1(\bX_i) \right)}{\sum_{i} \hat{e}(\bX_i) (1-\hat{e}(\bX_i))} \nonumber\\
&&-\frac{\sum_{i}\hat{e}(\bX_i) \left((1-Z_i) Y_i + (Z_i-\hat{e}(\bX_i))\hat{\mu}_0(\bX_i) \right)}{\sum_{i} (1-\hat{e}(\bX_i)) \hat{e}(\bX_i)}.
\end{eqnarray}
In the Supplementary Material A, we prove that $\hat{\tau}_{aug}^{\ATO}$ is consistent if the propensity score is correctly specified irrespective of the specification of the outcome model, but is not consistent if the propensity score is misspecified. Nonetheless, recent theoretical and empirical evidence suggests that overlap weighting is less sensitive to misspecification of the propensity scores than ATT and inverse probability weights \citep{zhou2020propensity}. This is because the region with the most overlap is subject to the least uncertainty in estimating the treatment effects. Moreover, it is known that the outcome model usually exerts a stronger influence on the augmented estimator than the propensity score model \cite[e.g.][]{li2013propensity}. In the case of ATO, this allows us to use a more stable outcome model to offset the potentially subjective propensity score model. Another advantage of the augmented estimators is that with correctly specified outcome model they are generally more efficient than their weighting counterparts \citep{LuncefordDavidian2004}, which is particularly desirable in the RD setting given the small sample size. 
}

We also derive M-estimation sandwich variance estimators \citep{vanderVaart1998,StefanskiBoos2002} for the augmented estimators  \eqref{eq:ATO_aug} and \eqref{eq:ATT_aug}, which account for the uncertainty in estimating the propensities from the ordered probit model \eqref{eq::probit_cut}. Exact forms of the estimators and details of derivation are given in the Supplementary Material B. {
However, these variance estimators do not take into account the uncertainty due to the choice of the bandwidth $d$ and thus may underestimate the true variance in practice. Note that this is a common problem in RD analysis where the bandwidth selection and inference on treatment effects is usually conducted in two separate steps; an integrated approach such as a full Bayesian model would avoid such a problem but is beyond the scope of this paper.} 


\subsubsection{Select the subpopulation} \label{sec:Select}
An important issue in practice is how to select the subpopulation $\Omega_0$ where Assumption \ref{as::unconfound} holds. There can be many choices of the shape of the subpopulation. 
We first focus on the symmetric intervals around the threshold: $(0.5-d)<\hat{e}(\bx_{i})<(0.5+d)$. To select the bandwidth $d$, we adopt the idea of balancing tests \citep{CattaneoEtAl2015, CattaneoVazquez-Bare2016}.
The basic idea is that, given the ``local'' nature of Assumption \ref{as::unconfound}, we expect the pretreatment covariates to be balanced between treatment groups close to the threshold, but the balance will break down when moving away from the threshold. Therefore, starting from a small $d$, we check the covariate balance of units in the interval $(0.5-d)<\hat{e}(\bx_{i})<(0.5+d)$, and gradually increase $d$ until significant imbalance is detected. 
Then, starting from the maximum $d$ such that the covariates are balanced, we make the interval asymmetric by increasing its length on the right or left. We continue doing so as long as all the covariates are balanced.
This allows us to find covariate-balanced subsamples with a larger number of units. As a result, asymmetric intervals allow us to increase the external validity of our findings.


\section{Empirical application}\label{sec:Empirical application}
\subsection{More Background of the CSPP}
The purchases in the CSPP can occur both in the primary and the secondary market. In order to be \textit{eligible} for purchase under the CSPP, \textit{debt instruments issued} must satisfy the following conditions:
(i) have a remaining maturity between 6 months and 31 years at the time of purchase;
(ii) be denominated in euro;
(iii) have a minimum first-best credit assessment of at least rating of BBB- or equivalent (i.e., investment-grade) obtained from an external and independent credit assessment institution;
(iv) provide a yield to maturity, which can also be negative, above the deposit facility rate.

In addition, the \textit{bond issuer} has to comply with the following requirements:
(i) is a corporation established in the euro area;
(ii) is not a credit institution supervised under the Single Supervisory Mechanism;
(iii) does not have a parent undertaking that is also a credit institution;
(iv) is not an investment firm, an asset management vehicle or a national asset management fund created in order to support financial sector restructuring;
(v) has not issued an asset-backed security, a `multi cedula' or a structured covered bond;
(vi) must not have a parent company which is under banking supervision inside or outside the euro area, and must not be a subsidiary of a supervised entity or a supervised group;
(vii) is not an eligible issuer for the Public Sector Purchase Programme (PSPP).

We employ the methods proposed in Section \ref{sec:Methods} to evaluate the effects of the CSPP on bond spreads in the primary market. More specifically, we assess how the \emph{eligibility for purchase} under the CSPP affects bond spreads at the time of their issuance. We define the treatment as the eligibility for purchase rather than the actual purchase of the bond for the following reasons. First, purchases under the CSPP are not pre-announced, making it impossible for market participants to react to them. Second, given that most eligible bonds issued after the program was announced have been purchased by the Eurosystem, market participants are likely to take the eligibility for purchase into consideration when pricing a bond at its issuance. Indeed, of the 346 eligible bonds that we ultimately use in our analysis, more than 85 per cent had been purchased by the Eurosystem as of January 26, 2018. Finally, owing to the relatively low liquidity of the secondary bond market, the effect of the actual purchase can be expected to be highly bond-specific and potentially only short-lived. Any permanent effect of the program on spreads of eligible bonds is, instead, likely to be largely observed already at issuance. Defining the treatment in this manner, {and focusing on bonds that satisfy all the eligibility criteria of the program with the exception of that pertaining to ratings,} implies that its effect can be evaluated using a \textit{sharp} RD design.

Having defined the treatment as the eligibility for purchase, we classify all bonds whose highest rating is equal to or greater than BBB- as treated units and the remaining bonds as control units. This does not imply that the treatment is equivalent to being assigned an investment-grade rating, due to the fact that market participants employ either the average or the minimum rating to identify investment-grade bonds \citep{AbidiFlores2018}. Therefore, the threshold employed by market participants is above that defining eligibility for purchase under the CSPP. {Following the extensive literature on bond ratings, studying their determinants and the information they contain, we model the running variable defined in this way as a function of bond and issuers characteristics. The seminal work in this literature, \citet{Hickman1958}, illustrates how ratings correlate with bond characteristics. Subsequently, \citet{PogueSoldofsky1969} employ linear regressions to show that information from firms’ financial statements explains ratings to a significant extent. \citet{PinchesMingo1973, PinchesMingo1975} reach a similar conclusion by means of a multiple discriminant analysis. \citet{AngPatel1975} conduct a comparative study of rating models and their ability to predict financial distress. \citet{KaplanUrwitz1979} propose an alternative approach based on an ordered probit model. \citet{KaoWu1990} use a similar ordered probit model to obtain estimates of the default risk of bonds, which in turn is an explanatory variable in a model explaining their yields. \citet{BlumeEtAl1998} extend the ordered probit framework to a setting with a time dimension.}

\subsection{Data}\label{sec:Data}
We employ two sources of proprietary data. First, we obtained from Bloomberg all the corporate bonds satisfying the eligibility criteria of the program {referring to characteristics other than the rating of the bond}, and issued between March 10, 2016 and September 30, 2017. A total of 899 such bonds were found. {This sample is representative of the population of euro-denominated bonds issued by euro-area non-bank corporations; the other eligibility criteria eliminate only 2 per cent of the bonds of our interest. Thus, we can estimate the effect of the program in this population using a sharp RD design.} The choice of the start date is motivated by the fact that already when the program was announced it became known that only investment-grade bonds would be eligible for purchase. We consider bonds issued after the program was announced as we wish to focus on the primary market. This is motivated by the relatively low liquidity of the secondary corporate bond markets in Europe \citep{BiaisEtAl2006,GunduzEtAl2017}, rendering secondary market quotes noisy indicators of going prices. Primary market prices, on the contrary, provide accurate information about the market valuation of bonds at the time of their issuance.

For each bond, we obtained from Bloomberg the following information: International Securities Identification Number (ISIN), coupon rate (cpn), maturity type, issue date, original maturity (mat), amount sold, coupon type, rating at issuance by Standard \& Poor's, Moody's, Fitch and DBRS along with its option-adjusted spread. Maturity type refers to any embedded options the bond contains (callable, putable, convertible) or it being a bullet bond (at maturity). Coupon type is one of the following: fixed, zero-coupon, pay-in-kind or variable. The option-adjusted spread (OAS) compares the yield to maturity of the bond to the yield to maturity of a government bond with a similar maturity, and further accounts for any embedded option features of the bond. 
For the OAS, the first available value between the issue date and the subsequent eight days was employed. We also obtained from Bloomberg the country of incorporation and the industry (as given by the Bloomberg Industry Classification System) of the issuer of each bond. Due to the difficulty of comparing bonds with variable coupon rates to fixed rate bonds, we excluded the former (6 bonds) from the analysis. 


We illustrate, in Figure \ref{fig:OASbyRating}, how the option-adjusted spreads vary across bonds issued during the program with different ratings (right in each pair). For the sake of comparison, the distributions of the OAS are presented also for bonds issued before the announcement of the CSPP, between September 21, 2014 and March 9, 2016. This time interval was chosen as it is similar in length to that of the program data. For all rating categories, apart from the highest two, the option-adjusted spreads were lower during the program than before it. A particularly notable difference is observed for the lowest investment-grade category, BBB-.

\begin{figure}[tbh]
	\centering
	\includegraphics[scale=0.65]{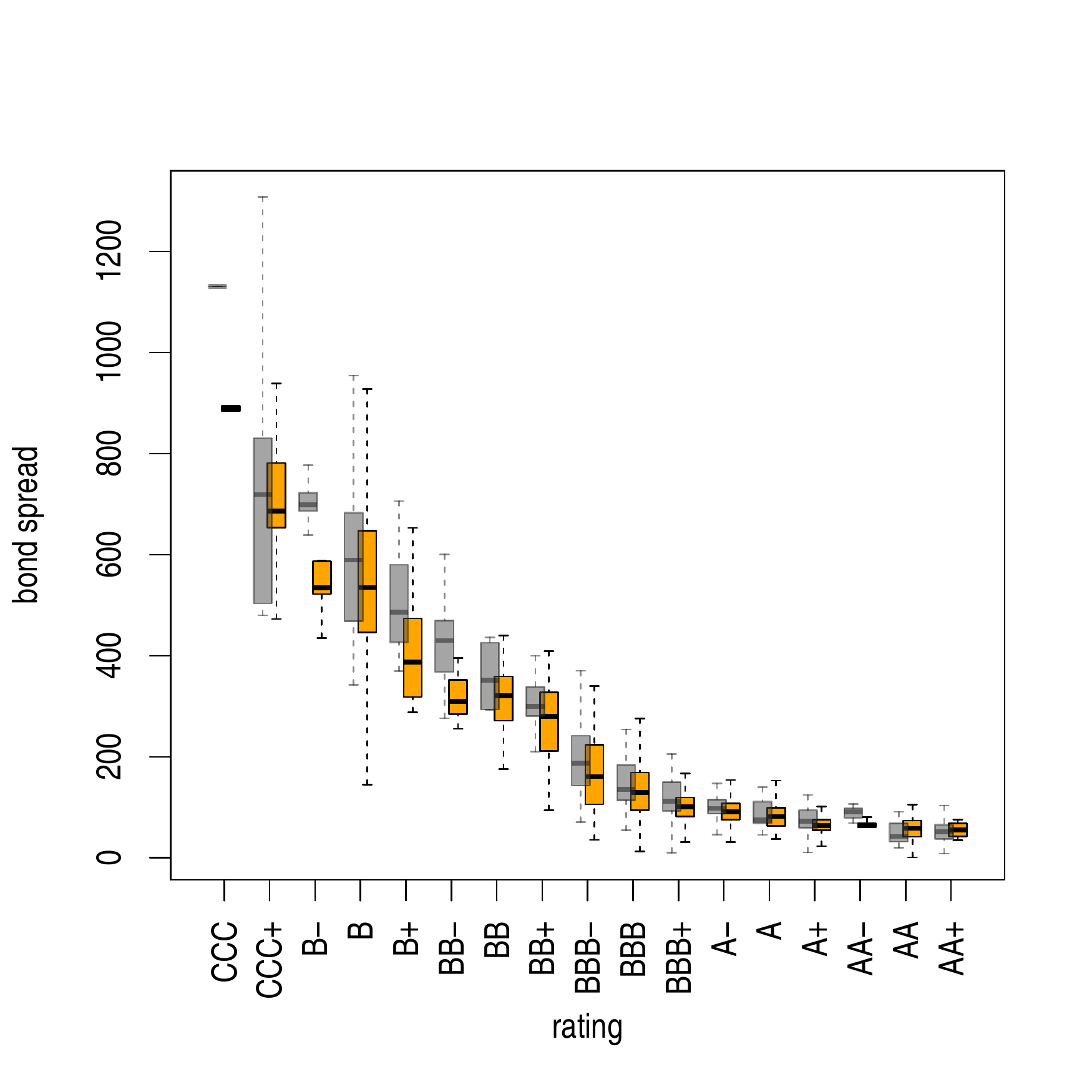}
	\caption{OAS by rating, before (left in each pair) and during the program.}
	\label{fig:OASbyRating}
\end{figure}

The second source of data employed is S\&P Capital IQ, from which we obtained balance sheet (BS) and income statement (IS) data for the bond issuers. More specifically, we first identified the ultimate parent company of each subsidiary issuer. Then, for these ultimate parents and the issuers with no parent companies, we obtained the following BS and IS items for the fiscal year 2015: earnings before interest and taxes (EBIT), total revenue, cash from operations, total assets, total liabilities, interest expenses, total debt, common equity and long-term debt. In addition, we recorded the year in which the company had been founded. When no data existed for the ultimate parent company, for instance due to it being a private company, we obtained data for the parent on the highest level in the corporate structure for which data was available. From the recorded data, we constructed the following variables: profitability (prof), cash flow (cf), liquidity (liq), interest coverage (cov), leverage (lev), solvency (solv), size, age and long-term debt (ltdebt). They are described in Table \ref{tab:SummaryStatsCiqData}.
We chose these variables as they are known to be determinants of credit quality \citep{BlumeEtAl1998,MizenTsoukas2012}. Units for which we obtained anomalous variable values, suggesting erroneously recorded BS or IS items, were excluded from the calculation of the summary statistics and the rest of the analysis. More specifically, we excluded the bonds issued by companies for which interest coverage exceeded 250 (3 companies), leverage exceeded 1 (3 companies) and solvency was below -1 (1 company). These exclusions led to the removal of 29 bonds.

\begin{table}[tbh]
	\caption{Summary statistics for the issuer characteristics.}
	\label{tab:SummaryStatsCiqData}
	\vspace{0.2em}
	\begin{center}
	{\renewcommand{\arraystretch}{1.1}
		\begin{tabular}{@{}l*{7}{c}@{}}
			\hline\hline
			variable& definition													& mean 	& sd 	& Q$_1$	& Q$_2$	& Q$_3$ & N  \\ \hline
			prof	& $\frac{\text{EBIT}}{\text{total revenue}}$					& 0.14	& 0.27	& 0.046	& 0.098	& 0.17	& 766\\		
			cf		& $\frac{\text{cash from operations}}{\text{total assets}}$		& 0.055	& 0.095	& 0.033	& 0.067	& 0.096	& 699\\
			liq		& $\frac{\text{cash from operations}}{\text{total liabilities}}$& 0.10	& 0.11	& 0.049	& 0.095	& 0.15	& 699\\
			cov		& $\frac{\text{EBIT}}{\text{interest expenses}}$				& 7.2	& 17	& 1.4	& 3.6	& 6.9	& 727\\
			lev		& $\frac{\text{total debt}}{\text{total assets}}$				& 0.37	& 0.20	& 0.24	& 0.35	& 0.49	& 746\\
			solv	& $\frac{\text{common equity}}{\text{total assets}}$			& 0.29	& 0.20	& 0.17	& 0.28	& 0.41	& 756\\
			size	& {\scriptsize $\log (\text{total revenue})$}					& 3.6	& 1.0	& 3.0	& 3.8	& 4.4	& 772\\
			age		& {\scriptsize 2017 \!$-$\! year founded}						& 77	& 76	& 22	& 61	& 115	& 709\\
			ltdebt	& $\frac{\text{long-term debt}}{\text{total assets}}$			& 0.33	& 0.35	& 0.16	& 0.26	& 0.40	& 747\\
			\hline
		\end{tabular}
	}
	\end{center}
	\begin{center}
	{\footnotesize NOTES: The variable size is calculated with total revenue recorded in millions of euros.}
	\end{center}
\end{table}

In the following analysis, we restrict attention to bonds for which data about their coupon rate, original maturity and all the characteristics of their issuers, i.e., the pretreatment variables, is available. There are 591 such bonds, of which 29 are convertible, 351 callable and 211 bullet bonds. However, the convertible bonds are not used in assessing the effect of the program as OAS is not available for them. We denote by \textit{call} the indicator variable equal to 1 if the bond is callable and 0 otherwise.

\subsection{Design}\label{sec:Design}
Our first objective is to obtain a well-specified ordered probit model for the running variable conditional on the pretreatment variables. In particular, we are concerned about how well the ordered probit model predicts ratings around the BBB- eligibility threshold. 
Relying on substantive knowledge, we first include the economically most relevant pretreatment variables that help predict ratings, leading to the following seven variables: cpn, mat, prof, cov, size, ltdebt and call. Then, we form all possible interaction and quadratic terms from these variables and include a combination of them that yields a model specification with adequate predictive power. {The final specification contains the predictors: cpn, mat, prof, cov, size, ltdebt, call, cpn$\times$size, cpn$\times$ltdebt, cpn$\times$call, mat$\times$prof, mat$\times$ltdebt, prof$\times$prof, prof$\times$call, cov$\times$call, size$\times$ltdebt, ltdebt$\times$ltdebt.}


To assess the goodness of fit of the model, we inspect how well it predicts the probability of being assigned to the treatment group. Figure \ref{fig:EstimatedPropensityScores} illustrates the distribution of the estimated propensity scores for each rating category. One observes that for high-yield bonds with a rating lower than BB and for investment-grade bonds with a rating higher than BBB the model predicts them to be with a high probability in the control and in the treatment group, respectively. Moreover, even for the four rating categories from BB to BBB around the threshold, the model correctly predicts the treatment status of most units. Specifically, the estimated propensity score is less than 0.5 for 60\% of the BB+ and BB bonds. For the lowest investment-grade categories BBB- and BBB, the estimated propensity score is greater than 0.5 for 98\% of the bonds in these two categories. Most importantly, Figure \ref{fig:EstimatedPropensityScores} shows that all the bonds with estimated propensity scores around 0.5 have ratings that are close to the investment grade threshold BBB-, suggesting the probit model is well specified.

\begin{figure}[tbh]
	\caption{Estimated propensity scores by rating.}
	\begin{center}
	\includegraphics[scale=0.6,trim={0 1cm 0 2cm}]{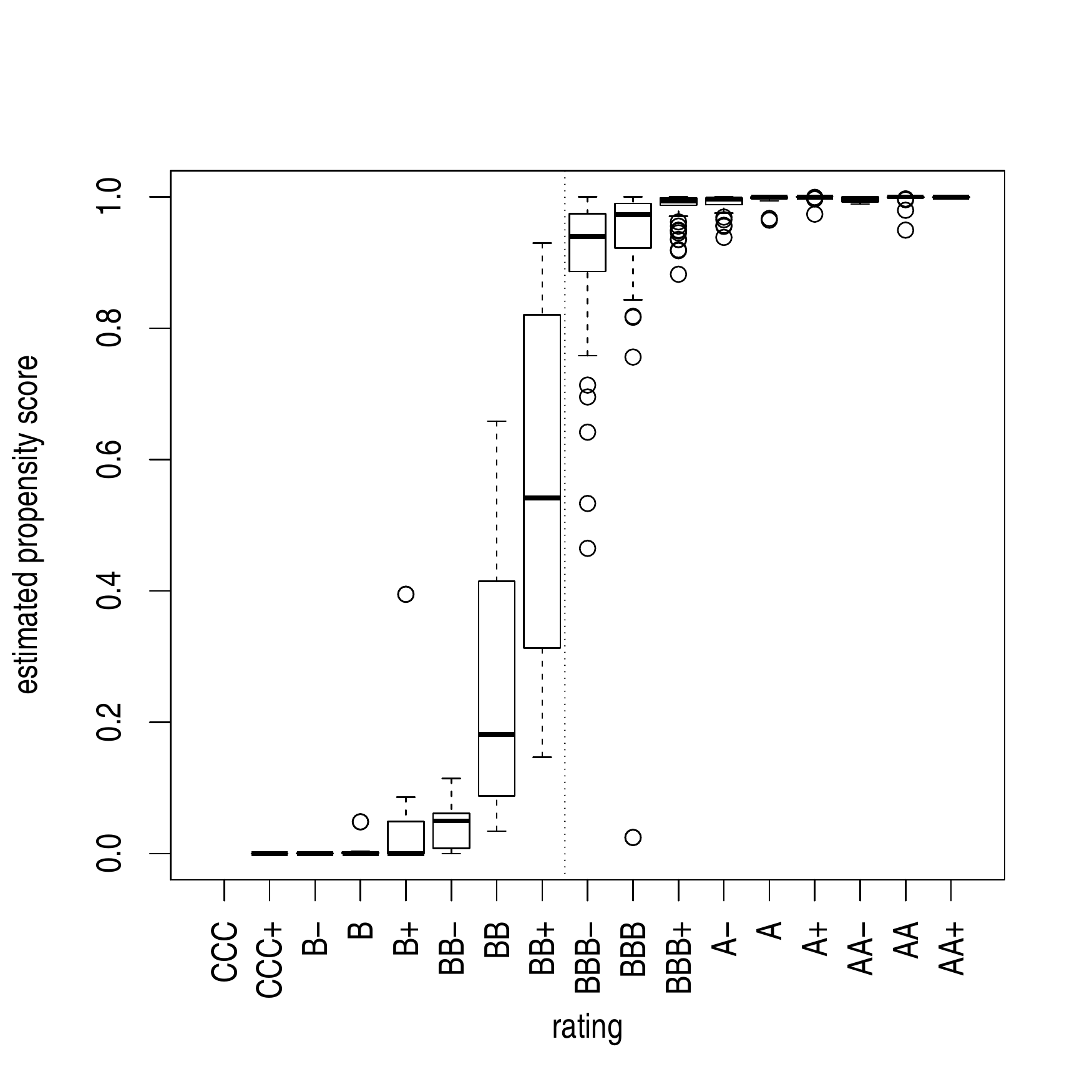}
	\end{center}
	\label{fig:EstimatedPropensityScores}
\end{figure}

Our second objective is to identify subsamples in which the distributions of the covariates are balanced between the treatment and control groups. Following the procedure in Section \ref{sec:Select}, we first construct subsets of units in which the estimated propensity score of each unit falls in the interval $(0.5-d,0.5+d)$, for some $d$. Then, in each subsample and for each pretreatment variable, we assess balance as by the standardized bias (SB):
\begin{equation*}
\text{SB} = \left( \frac{\sum_{i=1}^n X_i Z_i w_i}{\sum_{i=1}^n Z_i w_i} - \frac{\sum_{i=1}^n X_i ( 1 - Z_i ) w_i}{\sum_{i=1}^n ( 1 - Z_i ) w_i} \right) \bigg/ \sqrt{s_0^2/n_0 + s_1^2/n_1},
\end{equation*}
where $s_z^2$ is the sample variance of the unweighted covariate and $n_z$ the sample size in group $z=0,1$. When each unit is assigned a weight of unity, the SB is simply the two-sample $t$-statistic. 
We first calculate the SB using the overlap weights. Our goal is to find subsamples in which all the covariates are well balanced, ensuring, at the same time, that the number of units in them is not too small. We identify five such values of $d$, and present the corresponding SBs of the covariates in Panel A of Table \ref{tab:BalancingTestsSymmetricIntervals}. All of the absolute values of the SBs are smaller than 1.96 (the critical value of the two-sample $t$-statistic at 0.05 level), suggesting overall satisfactory covariate balance between the treatment and control groups. This supports the plausibility of local unconfoundedness in the subsamples under consideration. 

\begin{table}[tbh]
	\caption{Standardized bias of the covariates when $0.5-d<\hat{e}(\bx_i)<0.5+d$.}
	\label{tab:BalancingTestsSymmetricIntervals}
	\vspace{0.2em}
	\centering
	{\setlength{\tabcolsep}{3.5pt}
		{\scriptsize
			\begin{tabular}{@{}lcD{.}{.}{2.2}*{2}{D{.}{.}{1.2}}D{.}{.}{2.2}*{2}{D{.}{.}{1.2}}D{.}{.}{2.2}D{.}{.}{1.2}*{4}{D{.}{.}{2.2}}@{}}
				\hline\hline
				$d$ & n & \multicolumn{1}{c}{cpn} & \multicolumn{1}{c}{mat} & \multicolumn{1}{c}{prof} & \multicolumn{1}{c}{cf} & \multicolumn{1}{c}{liq} & \multicolumn{1}{c}{cov} & \multicolumn{1}{c}{lev} & \multicolumn{1}{c}{solv} & \multicolumn{1}{c}{size} & \multicolumn{1}{c}{age} & \multicolumn{1}{c}{ltdebt} & \multicolumn{1}{r}{call} \\ \hline
				\multicolumn{14}{c}{\textit{Panel A. Overlap (ATO) weights}} \\
				0.34 & 27 & -1.31 & 1.19 & 1.06 & 0.97 & 1.51 & 0.52 & -0.28 & 1.26 & -1.89 & -1.76 & -0.34 & -0.43 \\
				0.35 & 28 & -1.43 & 1.38 & 0.75 & 0.81 & 1.45 & 0.37 & -0.34 & 1.40 & -1.74 & -1.39 & -0.45 & -0.40 \\
				0.36 & 32 & -1.59 & 1.65 & 0.83 & 0.62 & 1.40 & 0.52 & -0.69 & 1.68 & -1.66 & -1.17 & -0.80 & -0.38 \\
				0.37 & 33 & -1.76 & 1.74 & 0.85 & 0.38 & 1.18 & 0.45 & -0.17 & 1.34 & -1.72 & -1.31 & -0.23 & -0.36 \\
				0.38 & 36 & -1.29 & 1.84 & 1.03 & 0.72 & 1.54 & 0.40 &  0.12 & 1.75 & -1.89 & -1.07 &  0.01 & -0.40 \\
				\multicolumn{14}{c}{\textit{Panel B. ATT weights}} \\
				0.34 & 27 & -1.40 & 1.00 & 1.03 &  0.57 & 1.42 & 0.50 & -0.08 & 1.39 & -1.80 & -1.53 & -0.35 & -0.57 \\
				0.35 & 28 & -1.50 & 1.46 & 0.54 &  0.33 & 1.26 & 0.21 & -0.11 & 1.54 & -1.56 & -1.02 & -0.46 & -0.48 \\
				0.36 & 32 & -1.51 & 2.09 & 0.54 &  0.01 & 1.07 & 0.33 & -0.47 & 1.81 & -1.33 & -0.69 & -0.79 & -0.33 \\
				0.37 & 33 & -1.83 & 2.17 & 1.15 & -0.35 & 0.70 & 0.15 &  0.32 & 1.23 & -1.72 & -1.01 &  0.13 & -0.25 \\
				0.38 & 36 & -0.39 & 2.18 & 1.46 &  0.75 & 1.69 & 0.24 &  0.99 & 2.14 & -1.92 & -0.30 &  0.71 & -0.45 \\
				\multicolumn{14}{c}{\textit{Panel C. No weights}} \\
				0.34 & 27 & -2.37 & 0.69 & 0.75 &  0.35 & 1.40 & 0.92 & -1.37 & 2.12 & -0.96 & -0.85 & -1.63 & -0.97 \\
				0.35 & 28 & -2.52 & 1.19 & 0.28 &  0.10 & 1.24 & 0.65 & -1.47 & 2.31 & -0.70 & -0.36 & -1.80 & -0.91 \\
				0.36 & 32 & -2.76 & 1.78 & 0.43 & -0.22 & 1.09 & 0.92 & -2.10 & 2.72 & -0.51 &  0.17 & -2.43 & -0.87 \\
				0.37 & 33 & -3.05 & 1.83 & 1.10 & -0.58 & 0.73 & 0.77 & -0.92 & 2.07 & -1.00 & -0.14 & -1.08 & -0.82 \\
				0.38 & 36 & -2.95 & 1.81 & 1.27 & -0.57 & 0.80 & 0.51 & -0.74 & 2.46 & -1.10 &  0.10 & -0.94 & -0.76 \\
				\hline
			\end{tabular}
		}
	}
\end{table}


Panel B of Table \ref{tab:BalancingTestsSymmetricIntervals} presents the SB based on the ATT weights for each covariate in each of the identified subsamples. Here, the covariates remain balanced in the two subsamples with the fewest units. However, for the subsamples defined by $d > 0.37$, the absolute values of SBs of two covariates exceed 1.96. 
For comparison, Panel C in Table \ref{tab:BalancingTestsSymmetricIntervals} presents the SBs of the covariates in the unweighted subsamples. Significant imbalance is observed in some of the covariates. Specifically, the $t$-statistics for \emph{cpn} and \emph{solv} exceed 1.96 in all the five subsamples. Taken together, the results suggest that both the overlap and ATT weights improve the overall covariate balance, even though not for each individual covariate. The greatest improvement is observed when employing the overlap weights. 

Finally, we investigate whether covariate-balanced subsamples with a larger number of units can be found by rendering asymmetric the intervals in which the estimated propensity scores are required to lie. Specifically, for both ATO and ATT schemes, we first identify from Table \ref{tab:BalancingTestsSymmetricIntervals} the largest value of $d$ for which all the absolute SBs are smaller than 1.96. Then, starting from these symmetric intervals ($d=0.38$  for ATO and $d=0.35$ for ATT), we gradually increase the length of the interval on the right or left of 0.5 until significant imbalance emerges. 
In both weighting schemes, we are able to identify subsamples with a significantly larger number of units, allowing us to more precisely estimate the effect of the program, as well as to improve the external validity of our results. {The SBs of the covariates in these subsamples are reported in the Supplementary Material C.}

\subsection{Results}
Having identified the subsamples in which local unconfoundedness plausibly holds, we proceed to estimate the treatment effects. 
{We use the augmented estimators $\hat{\tau}_{aug}^{\ATO}$ in \eqref{eq:ATO_aug} and $\hat{\tau}_{aug}^{\ATT}$ in \eqref{eq:ATT_aug} for point estimates, with the same outcome model for both ATO and ATT:
$$\bE(Y|\bX)=\gamma_0 + \gamma_1 cpn + \gamma_2 cpn^2 + \gamma_3 mat + \gamma_4 mat^2 + \gamma_5 lev + \gamma_6 lev^2 + \gamma_7 call.$$
The above model is postulated for both treatment and control groups, which is plausible within the subpopulation $\Omega_{o}$ where the treatment is assumed to be randomized. However, the coefficients are estimated separately for the two groups of units. The outcome model is firmly rooted in economic theory: corporate bond spreads are determined by the terms of the issues (coupon rate, maturity and call provisions) and the probability of default of their issuers \citep{Merton1974}. The latter is in turn increasing in financial leverage \citep{KrausLitzenberger1973}. Standard errors were obtained using the M-estimators developed in the Supplementary Material B.}

\begin{table}[tbh]
	\caption{Estimates of the treatment effect based on the augmented weighting estimators, with subpopulation identified using  asymmetric intervals.}
	\label{tab:TreatmentEffectEstimatesAsymmetricIntervals}
	\vspace{0.2em}
	\centering
	\begin{tabular}{@{}l*{4}{c}r@{}}
	\hline \hline
			$\hat{e}_{\min}$ & $\hat{e}_{\max}$ & $n_0$ & $n_1$ & estimate & se ($p$-val.) \\ \hline
		\multicolumn{6}{c}{\textit{Panel A. ATO }} \\
		0.10 & 0.88 & 21 & 16 & $-$19.2 &  6.74 (0.004) \\
		0.09 & 0.88 & 23 & 16 & $-$17.8 &  6.99 (0.011) \\
		0.08 & 0.88 & 24 & 16 & $-$18.1 &  7.42 (0.015) \\
		0.07 & 0.88 & 27 & 16 & $-$17.4 &  7.53 (0.021) \\
		\multicolumn{6}{c}{\textit{Panel B. ATT}} \\
		0.12 & 0.85 & 19 & 11 & $-$21.4 & 11.9 (0.073) \\
		0.11 & 0.85 & 20 & 11 & $-$21.6 & 11.9 (0.069) \\
		0.09 & 0.85 & 22 & 11 & $-$20.8 & 11.7 (0.075) \\
		0.08 & 0.85 & 23 & 11 & $-$21.1 & 11.6 (0.069) \\
		0.07 & 0.85 & 26 & 11 & $-$21.1 & 11.5 (0.066) \\
		\hline
	\end{tabular}
\end{table}


Table \ref{tab:TreatmentEffectEstimatesAsymmetricIntervals} reports the estimated ATO (Panel A) and ATT (Panel B) in the covariate-balanced subsamples (asymmetric intervals) identified in Section \ref{sec:Design}. The results based on the symmetric intervals are similar (and thus are not shown here), with slightly larger standard errors because of the smaller sample sizes. The results suggest that eligibility for purchase under the CSPP had a statistically significant and negative effect on bond spreads at issuance: a reduction in the range of 17-22 basis points. The magnitude of the ATT estimates is slightly larger than the ATO, but overall the results are consistent  across the subpopulations and weighting schemes. Our results are lower than the 70 basis point reduction in the primary market found by \citet{Zaghini2019}. However, the difference could simply reflect the more ``local'' nature of our estimates compared to those in \citet{Zaghini2019}, which are based on all the bonds issued in the primary market. Relative to the announcement effect of the program in the secondary bond market, our estimates are similar to the 15 basis point decrease reported in \citet{AbidiFlores2018}.   


Given the weighted average maturity of 7.8 years in the subsample defined by $\hat{e}_{\min}=0.07$ and $\hat{e}_{\max}=0.88$, 17--22 basis point reduction in yield to maturity corresponds approximately to a 1.3--1.7 per cent increase in the price of a zero-coupon bond at issuance.  Relative to the weighted average amount sold of 570 million euros in the subsample under consideration, this represents a significant decrease in the funding costs faced by the issuers of the eligible bonds.

As the effect of the program on bond spreads at issuance could have been due to higher expected liquidity of the eligible bonds, it is instructive to compare the effect that we have estimated to liquidity premia of corporate bonds. \citet{Dick-NielsenEtAl2012} estimate the liquidity premia of BBB US corporate bonds to lie in the range of 4--93 basis points. Also relative to these additional yields required by investors to compensate for the illiquidity of corporate bonds, our estimates of the effect of the program are {non-negligible}.

{
\subsection{A falsification test using negative controls} \label{sec:falsification}
It is possible that our analysis captures not only the causal effect of the CSPP but also the effect related to (1) the monetary policies that were already in place prior to the CSPP and (2) the rating of the bond. To examine this possibility, we conduct a falsification test using negative controls \citep{Rosenbaum2002}. Specifically, we perform a ``no treatment'' evaluation by re-running the analysis on a sample of corporate bonds issued in the pre-CSPP period. This pre-program sample contains all the corporate bonds issued between September 21, 2014 and March 9, 2016 that satisfy the eligibility criteria of the program with the exception of that pertaining to ratings. This sample is similar in terms of the length of the time interval, sample size, and unit characteristics to that used in the main analysis. More information about this sample is given in the Supplementary Material D. Given that the CSPP was not in place, we expect to observe no effect of a ``false'' treatment defined exactly as the real treatment (i.e.\ the highest bond rating is greater or equal to BBB-) in the main analysis.

The ATO and ATT estimates of the pre-CSPP sample are provided in Table \ref{tab:results_NC}. These estimates are negative but are not statistically significantly different from zero at the 10 \% level (p-value of 0.138 for ATO and 0.119 for ATT). This bolsters the conclusion that the treatment effects obtained in the main analysis are primarily attributed to the CSPP rather than other monetary policy measures or the rating of the bond itself.

\begin{table}[tbh]
	\caption{Effects of a ``false" treatment in the pre-CSPP period}
	\label{tab:results_NC}
	\vspace{0.2em}
	\centering
	\begin{tabular}{@{}l*{4}{c}r@{}}
		\hline \hline
		$\hat{e}_{\min}$ & $\hat{e}_{\max}$ & n$_0$ & n$_1$ & estimate & se ($p$-val.) \\ \hline
		\multicolumn{6}{c}{\textit{Panel A. ATO }} \\
		0.04 & 0.75 & 25 & 14 & $-$29.1 & 19.6 (0.138) \\ 		
		\multicolumn{6}{c}{\textit{Panel B. ATT}} \\
		0.01 & 0.73 & 33 & 11 & $-$31.0 & 19.9 (0.119) \\
		\hline
	\end{tabular}
\end{table}
}

\subsection{Alternative approaches}
An influential alternative approach is due to \citet{AngristRokkanen2015} (AR hereafter), who propose to identify causal effects away from the threshold by relying on a conditional independence assumption (CIA). In particular, they assume, conditional on a set of predictors which does not contain the running variable, the potential outcomes are mean-independent of the running variable. The CIA is similar to our local unconfoundedness assumption, and thus we also estimate the effect of being eligible for purchase under the CSPP employing the framework of AR. 
Here we invoke the bounded version of the CIA (BCIA) that is more plausible in our application: there exists $d > 0$ such that $\mathbb{E}[Y_i(z) | R_i, \bX_i, |R_i - r_t| < d] = \mathbb{E}[Y_i(z) | \bX_i, |R_i - r_t| < d]$, $z=0,1$, meaning that conditional mean-independence holds in a $d$-neighborhood of the threshold.

AR propose to identify the $d$-neighborhood of the threshold based on the value of the running variable. However, the measure of distance in the definition of the BCIA is not directly applicable to ordinal running variables. Instead, we take advantage of the ordered nature of the running variable and identify the set of units around the threshold as those with a rating BB+ (the highest category in the control group, 43 units total) or BBB- (the lowest category in the treatment group, 26 units total). We assume that the BCIA holds in this subset of units. AR propose to assess the BCIA by testing the coefficients in regressions of the outcome on the running variable and the pretreatment variables on either side of the threshold. This procedure is, again, not applicable to our selected subsample because we are only considering one category on each side of the threshold. Instead, we can check the covariate balance within this subsample (i.e., all bonds with BB+ and BBB- ratings) as an indirect assessment of the BCIA. We found that the covariate distribution between the treated and the control groups is strongly imbalanced: notably, nearly all 12 covariates have larger SBs than in our proposed method, 4 of which are larger than 1.96. The imbalance casts doubts in the validity of the BCIA in our application. Nonetheless, if we applied the AR approach despite this imbalance, we would obtain statistically insignificant estimates of the effect of the program: -36.7 basis points (p-value: 0.367) when employing the ATT weights. 

Overall, this comparison highlights a strength of our approach: we define the candidate subsamples based on richer covariate information, encoded in the estimated propensity scores obtained from the ordered probit model, and this helps identify subsets of units with better covariate balance between the treated and the control groups. Covariate balance lends powerful support to the validity of local unconfoundedness, being a stronger consequence of this assumption than the regression-based independence assumption invoked by AR.

\section{Conclusion}\label{sec:Conclusions}

In this paper, we have developed a regression discontinuity design applicable when the running variable, determining assignment to treatment, is ordinal. The estimation strategy is based on the following steps. We first estimate an ordered probit model for the ordinal running variable conditional on pretreatment variables. The estimated probability of being assigned to treatment is then adopted as a continuous surrogate running variable. In order to provide external validity to the analysis, we move away from the standard inference \emph{at} the threshold by assuming local unconfoundedness of the treatment in an interval \emph{around} the surrogate threshold. Then, once this interval has been identified via an overlap weighted balancing assessment of the preprogram variables across treatments, an estimate of the effect of the program in the interval is obtained employing an augmented weighting estimator of the average treatment effect.

We have applied our methodology to estimate the causal effect of the European Central Bank's Corporate Sector Purchase Programme (CSPP) on corporate bond spreads. 
We have estimated the effect of the program in a subpopulation defined by the estimated conditional probability to be eligible for purchase. This subpopulation is composed of bonds that can be assigned with non-negligible probability to either eligibility status, and therefore are the most affected by, even small, changes in the program. Our results suggest that eligibility for purchase under the CSPP had a negative effect, in the order of 17--22 basis points, on bond spreads at issuance. This is similar to previous estimates of the announcement effect of the program on bonds traded in the secondary market \citep{AbidiFlores2018}. Given that in the sample which is used to conduct inference the average amount issued exceeded 550 million euros, the 17--22 basis point reduction in the yield to maturity corresponds to a non-negligible decrease in the funding costs of the eligible issuers.


There are several limitations to our work. First, though our probit specification appears to perform well in the empirical application, it may be improved by a more objective procedure for choosing it. Specifically, our approach may give rise to a trade-off between variance and bias. Namely, when the model for the ordinal variable provides a good in-sample fit, the estimated propensity scores of most units are close to either 0 or 1. Consequently, covariate-balanced subsamples, identified using the estimated propensity scores, are likely to have moderate sample sizes. This may lead to elevated standard errors of the estimates of the treatment effect. One direction for future research is to develop a cross-validation criterion based on an objective function that achieves the right balance between bias and variance. Another direction is to conduct some sensitivity analysis on the model specification. 
Second, our method relies partially on the local SUTVA assumption, which rules out interference between units as well as any ``externality effects''. One could borrow from the recent advances to tackle the interference problem in causal inference to relax this assumption.

\section*{Supplement A:
Consistency properties of the augmented ATO estimator}
Recall the augmented weighting estimator of ATO, equation (3.6):
\begin{equation*}
\begin{aligned}
\hat{\tau}_{aug}^{ATO}
=&\frac{\sum_{i}^{}(1-\hat{e}_i) \left(Z_i Y_i-(Z_i-\hat{e}_i) \hat{\mu}_{1,i} \right)}{\sum_{i}^{} \hat{e}_i (1-\hat{e}_i)} \\
&-\frac{\sum_{i}^{}\hat{e}_i \left((1-Z_i) Y_i + (Z_i-\hat{e}_i)\hat{\mu}_{0,i} \right)}{\sum_{i}^{} (1-\hat{e}_i) \hat{e}_i},
\end{aligned}
\end{equation*}
where $e_i \equiv e(\bm{X}_i=\bm{x}_{i})=Pr(Z_i=1|\bX_i=\bm{x}_{i})$ is the propensity score, and $\mu_{z,i} \equiv \bE(Y_i(z_i)|\bX_i=\bm{x}_{i})$ is the regression model for the potential outcome $Y_i(z_i)$, for $z_i=0,1$.

Consistency requires:
\begin{equation*}
\hat{\tau}_z \rightarrow \Big[ \bE ( \ \underbrace{(1-e_i) \cdot e_i}_{h(\bm{x}_{i})} \cdot \mu_{z,i} ) \Big] \cdot \Big[ \bE ( \ \underbrace{(1-e_i) \cdot e_i}_{h(\bm{x}_{i})} ) \Big]^{-1},
\end{equation*}
for $z=0,1$.

Firstly, let us assume that both the propensity score $e_i$ and the regression outcome model $\mu_{z,i}$, $z=0,1$, are correctly specified, so that $\hat{e}_i \rightarrow e_i$, and $\hat{\mu}_{z,i} \rightarrow \mu_{z,i}$. Given the law of large numbers and the consistency of both $\hat{e}_i$ and $\hat{\mu}_{z,i}$, then $\hat{\tau}_1$ 
converges to $\bE \Big[ (1-e_i) \cdot e_i \cdot \mu_{1,i} \Big] \cdot \Big[ \bE \left((1-e_i) \cdot e_i \right) \Big]^{-1}$:

\begin{align*}
\hat{\tau}_1 =& \frac{\sum_{i}^{} (1-\hat{e}_i) Z_i Y_i - (1-\hat{e}_i) (Z_i-\hat{e}_i) \hat{\mu}_{1,i}} {\sum_{i}^{} \hat{e}_i (1-\hat{e}_i)} \\
\rightarrow &\Big[ \bE \left((1-e_i) Z_i \cdot Y_i(1) \right) - \bE\left( (1-e_i)(Z_i - e_i) \mu_{1,i} \right) \Big] \cdot \Big[ \bE \left(e_i (1-e_i) \right) \Big]^{-1}\\
=&\Big[ \bE\left( \bE \left((1-e_i) Z_i \cdot Y_i(1) \;\middle|\; \bX_i \right) \right) - \bE \Big( (1-e_i) \mu_{1,i} \underbrace{\bE((Z_i - e_i) |\bX_i)}_{0} \Big) \Big] \\
&\cdot \Big[ \bE \left(e_i (1-e_i) \right) \Big]^{-1} \\
=&\bE \Big[ (1-e_i)\bE(Z_i \cdot Y_i(1)|\bX_i) \Big] \cdot \Big[ \bE \left(e_i (1-e_i) \right) \Big]^{-1} \\
=&\bE \Big[ (1-e_i)\bE(Z_i|\bX_i) \cdot \bE (Y_i(1)|\bX_i)\Big] \cdot \Big[ \bE \left(e_i (1-e_i) \right) \Big]^{-1} \\
=&\bE \Big[ (1-e_i) \cdot e_i \cdot \mu_{1,i} \Big] \cdot \Big[ \bE \left((1-e_i) \cdot e_i \right) \Big]^{-1} 
\end{align*}

Similar arguments can be applied to prove that $\hat{\tau}_0$ achieves consistency. Moreover, since $(1-e_i) \mu_{1,i} \bE((Z_i - e_i) |\bX_i)$ is 
equal to zero irrespective of the specification for the regression outcome model, then the proof  holds also when the propensity score $e_i$ is correctly specified but $\mu_{z,i}$ is misspecified. Therefore $\hat{\tau}_{aug}^{ATO}$ is consistent when the propensity score is correctly specified irrespective of the correctness of the regression outcome model specification.

Now, let us assume that the propensity score is misspecified while the regression model is correctly specified, so that $\hat{e}_i \rightarrow \tilde{e}_i \neq e_i$, and $\hat{\mu}_{z,i} \rightarrow \mu_{z,i}$.
In this case consistency does not hold for $\hat{\tau}_1$ since:
\begin{align*}
\hat{\tau}_1 =& \frac{\sum_{i}^{} (1-\hat{e}_i) Z_i Y_i - (1-\hat{e}_i) (Z_i-\hat{e}_i) \hat{\mu}_{1,i}} {\sum_{i}^{} \hat{e}_i (1-\hat{e}_i)} \\
\rightarrow &\Big[ \bE \left( (1-\tilde{e}_i) \bE (Z_i \cdot Y_i(1) |\bX_i) \right) -  \bE \left( (1-\tilde{e}_i)\mu_{1,i}\bE(Z_i - \tilde{e}_i|\bX_i) \right) \Big] \\
& \cdot \Big[ \bE (\tilde{e}_i (1-\tilde{e}_i) ) \Big]^{-1} \\
=& \Big[ \bE \left( (1-\tilde{e}_i) e_i \cdot \bE (Y_i(1)|\bX_i) \right) - \bE \left( (1-\tilde{e}_i) e_i \cdot \mu_{1,i} \right)  +\bE \left( (1-\tilde{e}_i)\tilde{e}_i \cdot \mu_{1,i} \right) \Big] \\
&\cdot \Big[ \bE (\tilde{e}_i (1-\tilde{e}_i) ) \Big]^{-1} \\
=&\bE \Big[ (1-\tilde{e}_i) \cdot \tilde{e}_i \cdot \mu_{1,i} \Big] \cdot \Big[ \bE (\tilde{e}_i \cdot (1-\tilde{e}_i) ) \Big]^{-1}\\
\neq& \bE \Big[ (1-e_i) \cdot e_i \cdot \mu_{1,i} \Big] \cdot \Big[ \bE (e_i \cdot (1-{e}_i)) \Big]^{-1}
\end{align*}

Analogous arguments can be applied to prove that $\hat{\tau}_0$ is not consistent. Therefore, when the propensity score is misspecified, $\hat{\tau}_{aug}^{ATO}$ is 
not consistent.

\renewcommand{\theequation}{B.\arabic{equation}}
\setcounter{equation}{0}

\section*{Supplement B:
Derivation of the M-estimator of variance for the augmented ATO and ATT estimators}
In this supplement, we derive an M-estimation-based sandwich variance estimator of the augmented weighting estimators defined in Section 3.2, which accounts for the uncertainty in estimating the propensity score and the outcome models.

Recall that in Section 3.2.1, we postulate an ordered probit model for the distribution of the ordered running variable, $\Pr(R_{i}=r_{j}|\bX_{i}=\bx_i)$, and consequently for the propensity score $e_i \equiv e(\bX_{i})$.
Specifically, we assume that each unit's observed category $R_i$ is determined by a latent normally distributed variable $R_{i}^*$ as follows:
\begin{equation}\label{eq::probit}
R_{i}^*=\bX_{i}^{T} \bm{\beta} + \mathcal{E}_{i}, ~~~ \mathcal{E}_{i} \sim N(0,1), ~~~ i=1,2,\ldots,N 
\end{equation}
and
\begin{equation} \label{eq::probit_cut}
R_i=\left\{
\begin{array}{ll}
r_1,   &\mbox{if } R_{i}^* \leq u_{1}, \\
r_{j}, &\mbox{if } u_{j-1} < R_{i}^* \leq u_{j},\\
r_{J}, &\mbox{if }  R_{i}^* > u_{J-1},\\
\end{array}
\right.
\end{equation}
where $u_{j} \in \{u_{0}, u_{1}, \ldots , u_{J-1}, u_{J} \}$ is a series of cutoff points, with $u_{0} = -\infty$ and $u_{J} = \infty$. That is, $R_{i}$ falls in category $r_{j}$ when the latent variable $R_{i}^*$ falls in the interval between $u_{j-1}$ and $u_{j}$. Based on this ordered probit model:
\begin{equation*}
\Pr(R_{i}=r_{j}|\bX_{i}=\bx_i)=\Pr(u_{j-1} < R_{i}^* \leq u_{j})=\Phi(u_{j} - \bx_{i}^{T} \bm{\beta}) - \Phi(u_{j-1} - \bx_{i}^{T} \bm{\beta})
\end{equation*}

The log likelihood function of the ordered probit model \eqref{eq::probit}--\eqref{eq::probit_cut} is:
\begin{equation*}
l(\bm{\eta}) \equiv \sum_{i=1}^{N}l_{i}(\bm{\eta})=\sum_{i=1}^{N}\sum_{j=1}^{J}w_{ij}(\log[\Phi(u_{j}-\bm{x}_{i}^{T}\bm{\beta} )-\Phi(u_{j-1}-\bm{x}_{i}^{T}\bm{\beta} ) ]  ),
\end{equation*}
where $\bm{\eta}=(u_1, \ldots, u_{J-1}, \bm{\beta}^{T})^{T}$, $\bm{\beta}$ is a vector of parameters and $w_{ij}=\one(R_{i}=r_{j}),u_{0}=-\infty,u_{J}=\infty$. Maximum likelihood estimates of the model parameters can be obtained by setting to zero the score function:
\begin{align*}
0&=\bm{S}(\bm{\eta})\equiv \sum_{i=1}^{N}\bm{S}_{i}(\bm{\eta})\equiv \sum_{i=1}^{N}\frac{\partial}{\partial \bm{\eta}}l_{i}(\bm{\eta}) \\ 
&=\sum_{i=1}^{N}\left(\frac{\partial}{\partial u_{1} }l_{i}(\bm{\eta}),\cdots,\frac{\partial}{\partial u_{J-1} }l_{i}(\bm{\eta}),\frac{\partial}{\partial \bm{\beta}^{T} }l_{i}(\bm{\eta})\right)^{T},
\end{align*}
where $\bm{S}(\cdot)$ denotes the score function, i.e., the first-order derivative of the log likelihood function with respect to the parameters of interest.

The regression model for the potential outcome $Y_i(z_i)$ is:
\begin{equation}\label{eq::outcome}
\mu_{z,i} \equiv \mu_z(\bX_i) = \bE(Y_i(z_i)|\bX_i=\bm{x}_{i}) = \bm{x}_{i}^{T}\bm{\gamma}_{z}, ~~~ 
i=1,2,\ldots,N, 
\end{equation}
where 
$\bm{\gamma}_{z}$ is a vector of parameters.
To obtain maximum likelihood estimates of $\bm{\gamma}_{z}$, the score equations to be solved are:
\begin{equation*}
0=\bm{S}(\bm{\gamma}_{z})\equiv \sum_{i=1}^{N}\bm{S}_{i}(\bm{\gamma}_{z})\equiv \sum_{i=1}^{N}\frac{\partial}{\partial \bm{\gamma}_{z}}l_i(\bm{\gamma}_{z})
\end{equation*}

\subsection*{Variance estimator for the augmented estimator of ATT}
Recall the augmented weighting estimator of ATT, equation (3.5):
\begin{equation}\label{eq:AIPW_ATT_standard}
\hat{\tau}_{aug}^{\ATT}=\hat{\tau}_1-\hat{\tau}_0=\frac{\sum_{i=1}^{n} Y_i Z_i}{\sum_{i=1}^{n} Z_i}
- \frac{\sum_{i=1}^{n} \frac{Y_i (1-Z_i) \hat{e}_i + \hat{\mu}_{0,i}(Z_i-\hat{e}_i)}{1-\hat{e}_i}}{\sum_{i=1}^{n} Z_i}, 
\end{equation}
where $n<N$ denotes the sample size of the covariate-balanced subsample around the threshold.

Firstly, focus on $\hat{\tau}_1=\frac{\sum_{i=1}^{n} Z_i Y_i}{\sum_{i=1}^{n} Z_i}$ in (\ref{eq:AIPW_ATT_standard}), from which we derive its estimating equation:
\begin{equation}\label{eq:estimating_tau1_1}
\sum_{i=1}^{n} U_i(\hat{\tau}_1)=\sum_{i=1}^{n} Z_i(Y_i-\hat{\tau}_1)=0,
\end{equation}
which can be framed in the context of M-estimation theory. Its first-order derivative with respect to $\tau_1$ is given by:
\begin{equation*}
\frac{\partial U_i(\tau_1)}{\partial \tau_1}=-Z_i
\end{equation*}

We expand the estimating equation for $\hat{\tau}_1$ around the true value $\tau_1$; this is a first-order Taylor series, and higher-order terms will be $o_p(1)$ under standard regularity conditions.
In fact, the mean value theorem applied to (\ref{eq:estimating_tau1_1}) yields:
\begin{equation*}\label{eq:taylor_tau1_1}
0=\frac{1}{\sqrt{n}}\sum_{i=1}^{n} U_i(\hat{\tau}_1)=\frac{1}{\sqrt{n}}\sum_{i=1}^{n} U_i(\tau_1)+\left(\frac{1}{n}\sum_{i=1}^{n} \frac{\partial U_i(\tau_1)}{\partial \tau_1} \bigg|_{\tau_1=\tilde{\tau}_1} \right)\sqrt{n}(\hat{\tau}_1-\tau_1),
\end{equation*}
where $\tilde{\tau}_1$ is between $\tau_1$ and $\hat{\tau}_1$. Since all the estimates are consistent, Slutsky's theorem can be applied. That is, the averages can be replaced by their expectations, and an $o_p(1)$ term added, yielding:
\begin{equation}\label{eq:taylor_tau1_2}
\sqrt{n}(\hat{\tau}_1-\tau_1)=\theta^{-1}\left(\frac{1}{\sqrt{n}}\sum_{i=1}^{n} U_i(\tau_1)\right)+o_p(1),
\end{equation}
where $\theta=\bE(Z_i)=\bE(h(\bX_i))=\bE(e_i)=P(Z_i=1)$.

Secondly, consider $\hat{\tau}_0$ in (\ref{eq:AIPW_ATT_standard}):
\begin{equation*}
\hat{\tau}_0=\frac{\sum_{i=1}^{n} \frac{Y_i (1-Z_i) \hat{e}_i + \hat{\mu}_{0,i}(Z_i-\hat{e}_i)}{1-\hat{e}_i}}{\sum_{i=1}^{n} Z_i}, 
\end{equation*}
from which we derive the estimating equation for $\hat{\tau}_0$:
\begin{equation}\label{eq:estimating_tau0_2}
\sum_{i=1}^{n} U_i(\hat{\tau}_0,\bm{\hat{\eta}},\bm{\hat{\gamma}}_0)=\sum_{i=1}^{n}\left(\frac{Y_i(1-Z_i) \hat{e}_i + \hat{\mu}_{0,i}(Z_i-\hat{e}_i)}{1-\hat{e}_i}-\hat{\tau}_0 Z_i \right)=0,
\end{equation}
where $\bm{\gamma}_0$ represents the parameters of the regression model for the potential outcome $Y_i(0)$, and $\bm{\eta}$ the vector of parameters of the ordered probit model. Equation (\ref{eq:estimating_tau0_2}) can also be framed in the context of M-estimation theory. Its gradient is given by:
\begin{align*}
\frac{\partial U_i(\tau_0,\bm{\eta},\bm{\gamma}_0)}{\partial \tau_0}&=-Z_i\\
\frac{\partial U_i(\tau_0,\bm{\eta},\bm{\gamma}_0)}{\partial \bm{\eta}}&=\frac{Y_i(1-Z_i)e_{\bm{\eta}}}{(1-e_i)^2} + \frac{\mu_{0,i}(Z_i-1)e_{\bm{\eta}}}{(1-e_i)^2}\\
\frac{\partial U_i(\tau_0,\bm{\eta},\bm{\gamma}_0)}{\partial \bm{\gamma}_0}&=\frac{(Z_i-e_i)}{1-e_i}\frac{\partial \mu_{0,i}}{\partial \bm{\gamma}_0},
\end{align*}
where $e_{\bm{\eta}} \equiv \frac{\partial e_i}{\partial \bm{\eta}}$.

We expand the estimating equation for $\hat{\tau}_0$ around the true values $(\tau_0,\bm{\eta},\bm{\gamma}_0)$, yielding:
\begin{align*}
0=&\frac{1}{\sqrt{n}}\sum_{i=1}^{n} U_i(\hat{\tau}_0,\bm{\hat{\eta}},\bm{\hat{\gamma}}_0)\\
=&\frac{1}{\sqrt{n}}\sum_{i=1}^{n} U_i(\tau_0,\bm{\eta},\bm{\gamma}_0)-\left(\frac{1}{n}\sum_{i=1}^{n} Z_i \right)\sqrt{n}(\hat{\tau}_0-\tau_0)\\
&+\left(\frac{1}{n}\sum_{i=1}^{n} \frac{(Y_i-\tilde{\mu}_{0,i})(1-Z_i)\tilde{e}_{\bm{\eta}}}{(1-\tilde{e}_i)^2} \right)^{T}\sqrt{n}(\hat{\bm{\eta}}-\bm{\eta})\\
&+\left(\frac{1}{n}\sum_{i=1}^{n} \frac{(Z_i-\tilde{e}_i)}{1-\tilde{e}_i}\frac{\partial \mu_{0,i}}{\partial \bm{\gamma}_0} \bigg|_{\bm{\gamma_0}=\tilde{\bm{\gamma}}_0} \right)^{T}\sqrt{n}(\hat{\bm{\gamma}}_0-\bm{\gamma}_0),\nonumber
\end{align*}
where $(\tilde{\tau}_0; \bm{\tilde{\eta}}; \bm{\tilde{\gamma}}_0)$ is in the line segment between $(\tau_0; \bm{\eta}; \bm{\gamma}_0)$ and $(\hat{\tau}_0; \bm{\hat{\eta}}; \bm{\hat{\gamma}}_0)$, $\tilde{e}_i \equiv e(\bX_i;\tilde{\bm{\eta}})$, $\tilde{\mu}_{0,i} \equiv \mu_0(\bX_i;\tilde{\bm{\gamma}}_0)$, and $\tilde{e}_{\bm{\eta}} \equiv \frac{\partial e_i}{\partial \bm{\eta}}\bigg|_{\bm{\eta}=\tilde{\bm{\eta}}}$.
Both $\bm{\eta}$ and $\bm{\gamma}_0$ are estimated from the larger sample of size $N$, while the latter equation refers to the smaller sample of size $n$.
However, following \cite{Randles1982}, the values $\hat{\bm{\eta}}$ and $\hat{\bm{\gamma}_0}$ can also be inserted in the above equation.

Since all the estimates are consistent, Slutsky's theorem can be applied, yielding:
\begin{equation}\label{eq:taylor_tau0_2}
\begin{aligned}
\sqrt{n}(\hat{\tau}_0-\tau_0)=&\theta^{-1}\left(\frac{1}{\sqrt{n}}\sum_{i=1}^{n} U_i(\tau_0,\bm{\eta},\bm{\gamma}_0)\right)\\
&+\theta^{-1}\bE\left(\frac{(Y_i-\mu_{0,i})(1-Z_i)e_{\bm{\eta}}}{(1-e_i)^2} \right)^{T}\sqrt{n}(\hat{\bm{\eta}}-\bm{\eta})\\
&+\theta^{-1}\bE\left(\frac{(Z_i-e_i)}{1-e_i}\frac{\partial \mu_{0,i}}{\partial \bm{\gamma}_0} \right)^{T}\sqrt{n}(\hat{\bm{\gamma}}_0-\bm{\gamma}_0)+o_p(1),
\end{aligned}
\end{equation}
where $\theta=\bE(Z_i)=\bE(h(\bX_i))=\bE(e_i)=P(Z_i=1)$.

Taking the difference between the two expansions (\ref{eq:taylor_tau1_2}) and (\ref{eq:taylor_tau0_2}), we get:
\begin{align*}
\sqrt{n}(\hat{\Delta}_{aug}^{ATT}-\Delta_{aug}^{ATT})&=\sqrt{n}(\hat{\tau}_1-\tau_1)-\sqrt{n}(\hat{\tau}_0-\tau_0)\\
&=\theta^{-1}\frac{1}{\sqrt{n}}\sum_{i=1}^{n} \mathcal{I}_i+o_p(1),
\end{align*}
which yields:
\begin{equation*}
\hat{\tau}_1-\hat{\tau}_0=(\tau_1-\tau_0)+(n\theta)^{-1}\sum_{i=1}^{n} \mathcal{I}_i+o_p(1),
\end{equation*}
where
\begin{equation}\label{eq:differ_tau1_tau0_2}
\mathcal{I}_i=\left(U_i(\tau_1)-U_i(\tau_0,\bm{\eta},\bm{\gamma}_0)\right)-\bm{H}^{T}_{\eta}\bm{E}^{-1}_{\eta\eta}\bm{S}_i(\bm{\eta})-\bm{H}^{T}_{\gamma_{0}}\bm{E}^{-1}_{\gamma_{0}\gamma_{0}}\bm{S}_i(\bm{\gamma}_0)
\end{equation}
and
\begin{align*}
\bm{H}_{\eta}&=\bE\left(\frac{(Y_i-\mu_{0,i})(1-Z_i)e_{\bm{\eta}}}{(1-e_i)^2} \right)\\
\bm{H}_{\gamma_{0}}&=\bE\left(\frac{(Z_i-e_i)}{1-e_i}{\frac{\partial \mu_{0,i}}{\partial \bm{\gamma}_0}} \right)\\
\sqrt{n}(\hat{\bm{\eta}}-\bm{\eta})&=\bm{E}^{-1}_{\eta\eta}\frac{1}{\sqrt{n}}\sum_{i=1}^{n}\bm{S}_i(\bm{\eta})+o_p(1)\\
\sqrt{n}(\hat{\bm{\gamma}}_0-\bm{\gamma}_0)&=\bm{E}^{-1}_{\gamma_{0}\gamma_{0}}\frac{1}{\sqrt{n}}\sum_{i=1}^{n}\bm{S}_i(\bm{\gamma}_{0})+o_p(1),
\end{align*}
with $\bm{S}_i(\bm{\eta})$ and $\bm{S}_i(\bm{\gamma}_{0})$ denoting the score function for the propensity score and for the outcome model, respectively, while $\bm{E}_{\eta\eta}=\bE\left(\frac{\partial^2}{\partial \bm{\eta}\partial \bm{\eta}^{T}}l(\bm{\eta})\right)$ and $\bm{E}_{\gamma_{0}\gamma_{0}}=\bE\left(\frac{\partial^2}{\partial \bm{\gamma_0}\partial \bm{\gamma_0}^{T}} l(\bm{\gamma_{0}}) \right)$ are the information matrices for the propensity score and for the outcome model, respectively.

The empirical M-estimation variance of the moment estimator in \eqref{eq:AIPW_ATT_standard} can be calculated by considering the empirical counterpart of (\ref{eq:differ_tau1_tau0_2}):
\begin{equation}\label{eq:varMequation_ATT}
\bV(\hat{\Delta}_{aug}^{ATT})=\bV(\hat{\tau}_1-\hat{\tau}_0)=(n\hat{\theta})^{-2}\sum_{i=1}^{n} \hat{\mathcal{I}}_i^2
\end{equation}
Specifically, the quantities $U_i(\tau_1)$ and $U_i(\tau_0,\bm{\eta},\bm{\gamma}_0)$ can be consistently estimated by relying on (\ref{eq:estimating_tau1_1}) and (\ref{eq:estimating_tau0_2}).
Moreover, the parameter $\theta$ and the vectors $\bm{H}_{\eta}$ and $\bm{H}_{\gamma_{0}}$ can be consistently estimated by:
\begin{equation}\label{eq:varMequation_ATT_2}
\begin{aligned}
\hat{\theta}&= \frac{1}{n}\sum_{i=1}^{n} \hat{e}_i\\
\widehat{\bm{H}}_{\eta}&= \frac{1}{n}\sum_{i=1}^{n} \left(\frac{(Y_i-\hat{\mu}_{0,i})(1-Z_i)\hat{e}_{\bm{\eta}}}{(1-\hat{e}_i)^2} \right)\\
\widehat{\bm{H}}_{\gamma_{0}}&= \frac{1}{n} \sum_{i=1}^{n} \left(\frac{(Z_i-\hat{e}_i)}{1-\hat{e}_i}\frac{\partial \widehat\mu_{0,i}}{\partial \bm{\gamma}_0} \right)
\end{aligned}
\end{equation}
Consistent ML estimates of the inverse of the two information matrices $\bm{E}^{-1}_{\eta\eta}$ and $\bm{E}^{-1}_{\gamma_{0}\gamma_{0}}$ in (\ref{eq:differ_tau1_tau0_2}) can be obtained when estimating the ordered probit model and the outcome model.


\subsection*{Variance estimator for the augmented estimator of ATO}
Recall the augmented weighting estimator of ATO, equation (3.6):
\begin{equation}\label{eq:AIPW_ATO_STD}
\begin{aligned}
\hat{\tau}_{aug}^{ATO}=\hat{\tau}_1-\hat{\tau}_0=&\frac{\sum_{i=1}^{n}(1-\hat{e}_i) \left(Z_i Y_i-(Z_i-\hat{e}_i)\hat{\mu}_{1,i} \right)}{\sum_{i=1}^{n} \hat{e}_i (1-\hat{e}_i)} \\
&-\frac{\sum_{i=1}^{n}\hat{e}_i \left((1-Z_i) Y_i + (Z_i-\hat{e}_i)\hat{\mu}_{0,i} \right)}{\sum_{i=1}^{n} (1-\hat{e}_i) \hat{e}_i}
\end{aligned}
\end{equation}
where, as in (\ref{eq:AIPW_ATT_standard}), $n<N$ denotes the sample size of the covariate-balanced subsample around the threshold.

Firstly, focus on $\hat{\tau}_1$ in (\ref{eq:AIPW_ATO_STD}), from which we derive its estimating equation:
\begin{equation}\label{eq:estimating_tau1_1_ATO}
\sum_{i=1}^{n} U_i(\hat{\tau}_1,\hat{\bm{\eta}},\hat{\bm{\gamma}}_{1})=\sum_{i=1}^{n}(1-\hat{e}_i)\left(Z_i Y_i-(Z_i-\hat{e}_i)\hat{\mu}_{1,i}- \hat{e}_i \hat{\tau}_1 \right)=0
\end{equation}
Its gradient is given by:
\begin{align*}
\frac{\partial U_i(\tau_1,\bm{\eta},\bm{\gamma}_{1})}{\partial \tau_1}=&-e_i(1-e_i)\\
\frac{\partial U_i(\tau_1,\bm{\eta},\bm{\gamma}_{1})}{\partial \bm{\eta}}=&\left(Z_i (\mu_{1,i}-Y_i) + \mu_{1,i} (1-2 e_i) + \tau_1 (2 e_i - 1)\right) e_{\bm{\eta}} \\
\frac{\partial U_i(\tau_1,\bm{\eta},\bm{\gamma}_{1})}{\partial \bm{\gamma}_{1}}=&(e_i-1)(Z_i-e_i)\frac{\partial \mu_{1,i}}{\partial \bm{\gamma}_{1}}
\end{align*}

We expand the estimating equation for $\hat{\tau}_1$ around the true values $(\tau_1,\bm{\eta},\bm{\gamma}_1)$, yielding:
\begin{align*}
0=&\frac{1}{\sqrt{n}}\sum_{i=1}^{n} U_i(\hat{\tau}_1,\bm{\hat{\eta}},\bm{\hat{\gamma}}_1)\\
&\frac{1}{\sqrt{n}}\sum_{i=1}^{n} U_i(\tau_1,\bm{\eta},\bm{\gamma}_1)-\left(\frac{1}{n}\sum_{i=1}^{n} \tilde{e}_i (1-\tilde{e}_i) \right)\sqrt{n}(\hat{\tau}_1-\tau_1)\\
&+\Bigg(\frac{1}{n}\sum_{i=1}^{n} \left(Z_i (\tilde{\mu}_{1,i}-Y_i) + \tilde{\mu}_{1,i} (1-2 \tilde{e}_i) + \tilde{\tau}_1 (2 \tilde{e}_i - 1)\right) \tilde{e}_{\bm{\eta}}\Bigg)^{T}\sqrt{n}(\hat{\bm{\eta}}-\bm{\eta})\\
&+\Bigg(\frac{1}{n}\sum_{i=1}^{n} \Bigg((\tilde{e}_i-1)(Z_i - \tilde{e}_i)\frac{\partial \mu_{1,i}}{\partial \bm{\gamma}_{1}}\bigg|_{\bm{\gamma_1}=\tilde{\bm{\gamma_1}}}\Bigg)\Bigg)^{T}\sqrt{n}(\hat{\bm{\gamma}}_1-\bm{\gamma}_1),
\end{align*}
where $(\tilde{\tau}_1; \bm{\tilde{\eta}}; \bm{\tilde{\gamma}}_1)$ is in the line segment between $(\tau_1; \bm{\eta}; \bm{\gamma}_1)$ and $(\hat{\tau}_1;\bm{\hat{\eta}};\bm{\hat{\gamma}}_1)$, $\tilde{e}_i \equiv e(\bX_i;\tilde{\bm{\eta}})$ and $\tilde{\mu}_{1,i} \equiv \mu_1(\bX_i;\tilde{\bm{\gamma}_1})$ \citep{Randles1982}. Since all the estimates are consistent, Slutsky's theorem can be applied as before, yielding:
\begin{equation}\label{eq:taylor_tau1_3_ATO}
\begin{aligned}
&\sqrt{n}(\hat{\tau}_1-\tau_1) = \theta^{-1}\frac{1}{\sqrt{n}}\sum_{i=1}^{n} U_i(\tau_1,\bm{\eta},\bm{\gamma}_1)\\
&+\theta^{-1}\bE\left(\left(Z_i (\mu_{1,i}-Y_i) + \mu_{1,i} (1-2 e_i) + \tau_1 (2 e_i - 1)\right) e_{\bm{\eta}}\right)^{T}\sqrt{n}(\hat{\bm{\eta}}-\bm{\eta})\\
&+\theta^{-1}\bE\Bigg((e_i - 1)(Z_i - e_i)\frac{\partial \mu_{1,i}}{\partial \bm{\gamma}_{1}} \Bigg)^{T}\sqrt{n}(\hat{\bm{\gamma}}_1-\bm{\gamma}_1)+o_p(1),
\end{aligned}
\end{equation}
where $\theta=\bE\left( e_i(1-e_i)\right)$.

Secondly, focus on $\hat{\tau}_0$ in (\ref{eq:AIPW_ATO_STD}), from which we derive its estimating equation:
\begin{equation}\label{eq:estimating_tau0_1_ATO}
\sum_{i=1}^{n} U_i(\hat{\tau}_0,\bm{\hat{\eta}},\bm{\hat{\gamma}}_0)
=\sum_{i=1}^{n}\hat{e}_i\left((1-Z_i)Y_i+(Z_i-\hat{e}_i)\hat{\mu}_{0,i}-(1-\hat{e}_i)\hat{\tau}_0\right)=0
\end{equation}
Its gradient is given by:
\begin{align*}
\frac{\partial U_i(\tau_0,\bm{\eta},\bm{\gamma}_{0})}{\partial \tau_0}&=-e_i(1-e_i)\\
\frac{\partial U_i(\tau_0,\bm{\eta},\bm{\gamma}_{0})}{\partial \bm{\eta}}&=((1-Z_i) Y_i + \mu_{0,i} (Z_i - 2 e_i) - \tau_0 (1 - 2 e_i))e_{\bm{\eta}}\\
\frac{\partial U_i(\tau_0,\bm{\eta},\bm{\gamma}_{0})}{\partial \bm{\gamma}_{0}}&=e_i(Z_i-e_i)\frac{\partial \mu_{0,i}}{\partial \bm{\gamma}_{0}}
\end{align*}

We expand the estimating equation for $\hat{\tau}_0$ around the true values $(\tau_0,\bm{\eta},\bm{\gamma}_0)$:
\begin{align*}\label{eq:taylor_tau0_2_ATO}
0=&\frac{1}{\sqrt{n}}\sum_{i=1}^{n} U_i(\hat{\tau}_0,\bm{\hat{\eta}},\bm{\hat{\gamma}}_0)\\
&\frac{1}{\sqrt{n}}\sum_{i=1}^{n} U_i(\tau_0,\bm{\eta},\bm{\gamma}_0)-\left(\frac{1}{n}\sum_{i=1}^{n} \tilde{e_i}(1-\tilde{e_i}) \right)\sqrt{n}(\hat{\tau}_0-\tau_0)\\
&+\left(\frac{1}{n}\sum_{i=1}^{n}
((1-Z_i) Y_i + \tilde{\mu}_{0,i} (Z_i - 2 \tilde{e_i}) - \tilde{\tau_0} (1 - 2 \tilde{e_i})) \tilde{e}_{\bm{\eta}} \right)^{T}\sqrt{n}(\hat{\bm{\eta}}-\bm{\eta})\\
&+\left(\frac{1}{n}\sum_{i=1}^{n} \left( \tilde{e}_i(Z_i-\tilde{e}_i) \frac{\partial \mu_{0,i}}{\partial \bm{\gamma}_{0}}\bigg|_{\bm{\gamma_0}=\tilde{\bm{\gamma_0}}}\right) \right)^{T}\sqrt{n}(\hat{\bm{\gamma}}_0-\bm{\gamma}_0),
\end{align*}
where $(\tilde{\tau}_0; \bm{\tilde{\eta}}; \bm{\tilde{\gamma}}_0)$ is in the line segment between $(\tau_0; \bm{\eta}; \bm{\gamma}_0)$ and $(\hat{\tau}_0; \bm{\hat{\eta}}; \bm{\hat{\gamma}}_0)$, $\tilde{e}_i \equiv e(\bX_i;\tilde{\bm{\eta}})$, $\tilde{\mu}_{0,i} \equiv \mu_0(\bX_i;\tilde{\bm{\gamma}}_0)$, and $\tilde{e}_{\bm{\eta}} \equiv \frac{\partial e_i}{\partial \bm{\eta}}\bigg|_{\bm{\eta}=\tilde{\bm{\eta}}}$ \citep{Randles1982}.
Since all the estimates are consistent, Slutsky's theorem can again be applied:
\begin{equation}\label{eq:taylor_tau0_3_ATO}
\begin{aligned}
&\sqrt{n}(\hat{\tau}_0-\tau_0)=\theta^{-1}\frac{1}{\sqrt{n}}\sum_{i=1}^{n} U_i(\tau_0,\bm{\eta},\bm{\gamma}_0)\\
&+\theta^{-1}\bE\left(((1-Z_i) Y_i + \mu_{0,i} (Z_i - 2 e_i) - \tau_0 (1 - 2 e_i)) e_{\bm{\eta}} \right)^{T}\sqrt{n}(\hat{\bm{\eta}}-\bm{\eta})\\
&+\theta^{-1}\bE\left( e_i(Z_i-e_i) \frac{\partial \mu_{0,i}}{\partial \bm{\gamma}_{0}} \right)^{T}\sqrt{n}(\hat{\bm{\gamma}}_0-\bm{\gamma}_0)+o_p(1),\\
\end{aligned}
\end{equation}
where $\theta=\bE\left( e_i(1-e_i)\right)$.

Taking the difference between the two expansions (\ref{eq:taylor_tau1_3_ATO}) and (\ref{eq:taylor_tau0_3_ATO}), we get:
\begin{align*}\label{eq:Mequation_ATO}
\sqrt{n}(\hat{\Delta}_{aug}^{ATO}-\Delta_{aug}^{ATO})&=\sqrt{n}(\hat{\tau}_1-\tau_1)-\sqrt{n}(\hat{\tau}_0-\tau_0)\\
&=\theta^{-1}\frac{1}{\sqrt{n}}\sum_{i=1}^{n} \mathcal{I}_i+o_p(1),\nonumber
\end{align*}
which yields:
\begin{equation*}
\hat{\tau}_1-\hat{\tau}_0=(\tau_1-\tau_0)+(n\theta)^{-1}\sum_{i=1}^{n} \mathcal{I}_i+o_p(1),
\end{equation*}
where
\begin{equation}\label{eq:differ_tau1_tau0_2_ATO}
\begin{aligned}
\mathcal{I}_i=&\left(U_i(\tau_1,\bm{\eta},\bm{\gamma}_1)-U_i(\tau_0,\bm{\eta},\bm{\gamma}_0)\right)\\
&-\bm{H}^{T}_{\eta}\bm{E}^{-1}_{\eta\eta}\bm{S}_i(\bm{\eta})- \bm{H}^{T}_{\gamma_{1}}\bm{E}^{-1}_{\gamma_{1}\gamma_{1}}\bm{S}_i(\bm{\gamma}_1)-\bm{H}^{T}_{\gamma_{0}}\bm{E}^{-1}_{\gamma_{0}\gamma_{0}}\bm{S}_i(\bm{\gamma}_0)
\end{aligned}
\end{equation}
and
\begin{align*}
&\bm{H}_{\eta}= - \bE\left( ( \mu_{1,i}(1 - 2 e_i + Z_i) - \mu_{0,i} (Z_i - 2 e_i) + (2 e_i -1) (\tau_1 - \tau_0) - Y_i ) e_{\bm{\eta}} \right) \\
&\bm{H}_{\gamma_{0}}=\bE\left( e_i(Z_i-e_i)\frac{\partial \mu_{0,i}}{\partial \bm{\gamma}_{0}} \right)\\
&\bm{H}_{\gamma_{1}}=\bE\left( (1-e_i)(Z_i-e_i) \frac{\partial \mu_{1,i}}{\partial \bm{\gamma}_{1}} \right)\\
&\sqrt{n}(\hat{\bm{\eta}}-\bm{\eta})=\bm{E}^{-1}_{\eta\eta}\frac{1}{\sqrt{n}}\sum_{i=1}^{n}\bm{S}_i(\bm{\eta})+o_p(1)\\
&\sqrt{n}(\hat{\bm{\gamma}}_0-\bm{\gamma}_0)=\bm{E}^{-1}_{\gamma_{0}\gamma_{0}}\frac{1}{\sqrt{n}}\sum_{i=1}^{n}\bm{S}_i(\bm{\gamma}_{0})+o_p(1)\\
&\sqrt{n}(\hat{\bm{\gamma}}_1-\bm{\gamma}_1)=\bm{E}^{-1}_{\gamma_{1}\gamma_{1}}\frac{1}{\sqrt{n}}\sum_{i=1}^{n}\bm{S}_i(\bm{\gamma}_{1})+o_p(1),
\end{align*}
with $\bm{S}_i(\bm{\eta})$, $\bm{S}_i(\bm{\gamma}_{0})$ and $\bm{S}_i(\bm{\gamma}_{1})$ denoting the score functions for the propensity score and for the outcome model for $z=0,1$, respectively, while $\bm{E}_{\eta\eta}=\bE\left(\frac{\partial^2}{\partial \bm{\eta}\partial \bm{\eta}^{T}}l(\bm{\eta})\right)$, $\bm{E}_{\gamma_{0}\gamma_{0}}=\bE\left(\frac{\partial^2}{\partial \bm{\gamma_0}\partial \bm{\gamma_0}^{T}} l(\bm{\gamma_{0}}) \right)$, $\bm{E}_{\gamma_{1}\gamma_{1}}=\bE\left(\frac{\partial^2}{\partial \bm{\gamma_1}\partial \bm{\gamma_1}^{T}} l(\bm{\gamma_{1}}) \right)$ are the information matrices for the propensity score and for the outcome model for $z=0,1$, respectively.

The empirical M-estimation variance of the augmented estimator in \eqref{eq:AIPW_ATO_STD} can be calculated by considering the empirical counterpart of (\ref{eq:differ_tau1_tau0_2_ATO}):
\begin{equation}\label{eq:varMequation_ATO}
\bV(\hat{\Delta}_{aug}^{ATO})=\bV(\hat{\tau}_1-\hat{\tau}_0)=(n\hat{\theta})^{-2}\sum_{i=1}^{n} \hat{\mathcal{I}}_i^2
\end{equation}
Specifically, the quantities $U_i(\tau_1,\bm{\eta},\bm{\gamma}_0)$ and $U_i(\tau_0,\bm{\eta},\bm{\gamma}_0)$ can be consistently estimated by relying on (\ref{eq:estimating_tau1_1_ATO}) and (\ref{eq:estimating_tau0_1_ATO}). Moreover, the parameter $\theta$ and the vectors $\bm{H}_{\eta}$ and $\bm{H}_{\gamma_{0}}$, $\bm{H}_{\gamma_{1}}$ can be consistently estimated by:
\begin{equation}\label{eq:varMequation_ATO_2}
\begin{aligned}
\hat{\theta}=& \frac{1}{n}\sum_{i=1}^{n} \hat{e}_i(1-\hat{e}_i)\\
\widehat{\bm{H}}_{\eta}=& - \frac{1}{n}\sum_{i=1}^{n} \left(\hat{\mu}_{1,i}(1 - 2 \hat{e_i} + Z_i) - \hat{\mu}_{0,i} (Z_i - 2 \hat{e_i}) + (2 \hat{e_i} -1) (\hat{\tau}_1 - \hat{\tau}_0) - Y_i\right)\hat{e}_{\bm{\eta}}\\
\widehat{\bm{H}}_{\gamma_{0}}=& \frac{1}{n}\sum_{i=1}^{n} \left( \hat{e}_i(Z_i-\hat{e}_i)\frac{{\partial \widehat \mu_{0,i}}}{\partial \bm{\gamma}_{0}}\right)\\
\widehat{\bm{H}}_{\gamma_{1}}=& \frac{1}{n}\sum_{i=1}^{n} \left( (1-\hat{e}_i)(Z_i-\hat{e}_i) \frac{{\partial \widehat\mu_{1,i}}}{\partial \bm{\gamma}_{1}} \right)
\end{aligned}
\end{equation}

Consistent estimators of the inverse of the two information matrices $\bm{E}^{-1}_{\eta\eta}$ and $\bm{E}^{-1}_{\gamma_{0}\gamma_{0}}$, $\bm{E}^{-1}_{\gamma_{1}\gamma_{1}}$ in equation (\ref{eq:differ_tau1_tau0_2_ATO}) can be obtained when estimating the ordered probit model and the outcome model.

\renewcommand{\theequation}{C.\arabic{equation}}
\setcounter{equation}{0}
\renewcommand{\thetable}{C.\arabic{table}}
\setcounter{table}{0}

\section*{Supplement C:
Covariate balance in the subsamples defined by asymmetric intervals}

The SBs of the covariates in the subsamples identified in Section 4.3 which are defined by asymmetric intervals are presented in Table \ref{tab:BalancingTestsAsymmetricIntervals}.

\begin{table}[tbh]
	\caption{Standardized bias of the covariates when $\hat{e}_{\min}<\hat{e}(x_i)<\hat{e}_{\max}$.}
	\label{tab:BalancingTestsAsymmetricIntervals}
	\vspace{0.2em}
	\centering
	{\setlength{\tabcolsep}{3.5pt}
		{\scriptsize
			\begin{tabular}{@{}lccD{.}{.}{2.2}*{5}{D{.}{.}{1.2}}D{.}{.}{2.2}D{.}{.}{1.2}*{4}{D{.}{.}{2.2}}@{}}
				\hline\hline
				$\hat{e}_{\min}$ & $\hat{e}_{\max}$ & n & \multicolumn{1}{c}{cpn} & \multicolumn{1}{c}{mat} & \multicolumn{1}{c}{prof} & \multicolumn{1}{c}{cf} & \multicolumn{1}{c}{liq} & \multicolumn{1}{c}{cov} & \multicolumn{1}{c}{lev} & \multicolumn{1}{c}{solv} & \multicolumn{1}{c}{size} & \multicolumn{1}{c}{age} & \multicolumn{1}{c}{ltdebt} & \multicolumn{1}{r}{call} \\ \hline
				\multicolumn{15}{c}{\textit{Panel A. Overlap (ATO) weights}} \\
				0.10 & 0.88 & 37 & -1.36 & 1.85 & 1.05 & 0.72 & 1.55 & 0.43 & 0.07 & 1.77 & -1.84 & -1.04 & -0.03 & -0.42 \\
				0.09 & 0.88 & 39 & -1.48 & 1.84 & 1.05 & 0.67 & 1.36 & 0.33 & 0.10 & 1.61 & -1.88 & -0.95 & -0.01 & -0.44 \\
				0.08 & 0.88 & 40 & -1.54 & 1.82 & 1.05 & 0.67 & 1.37 & 0.35 & 0.06 & 1.65 & -1.88 & -0.97 & -0.05 & -0.45 \\
				0.07 & 0.88 & 43 & -1.72 & 1.84 & 1.08 & 0.74 & 1.43 & 0.39 & 0.08 & 1.64 & -1.87 & -1.00 & -0.03 & -0.48 \\				
				\multicolumn{15}{c}{\textit{Panel B. ATT weights}} \\
				0.12 & 0.85 & 30 & -1.61 & 1.44 & 0.58 & 0.33 & 1.29 & 0.24 & -0.19 & 1.65 & -1.63 & -0.99 & -0.55 & -0.50 \\
				0.11 & 0.85 & 31 & -1.65 & 1.45 & 0.60 & 0.33 & 1.29 & 0.26 & -0.23 & 1.68 & -1.60 & -0.97 & -0.59 & -0.51 \\
				0.09 & 0.85 & 33 & -1.69 & 1.44 & 0.61 & 0.31 & 1.19 & 0.22 & -0.22 & 1.59 & -1.63 & -0.92 & -0.61 & -0.53 \\
				0.08 & 0.85 & 34 & -1.73 & 1.44 & 0.61 & 0.31 & 1.20 & 0.23 & -0.25 & 1.62 & -1.64 & -0.93 & -0.64 & -0.54 \\
				0.07 & 0.85 & 37 & -1.84 & 1.45 & 0.65 & 0.34 & 1.24 & 0.25 & -0.25 & 1.65 & -1.64 & -0.93 & -0.66 & -0.56 \\	
				\hline
			\end{tabular}
		}
	}
\end{table}


\renewcommand{\theequation}{D.\arabic{equation}}
\setcounter{equation}{0}
\renewcommand{\thetable}{D.\arabic{table}}
\setcounter{table}{0}
\renewcommand{\thefigure}{D.\arabic{figure}}
\setcounter{figure}{0}

\section*{Supplement D:
Information about the pre-CSPP sample in the falsification test}

The pre-program sample used in the falsification test in Section 4.5 contains all the corporate bonds issued between September 21, 2014 and March 9, 2016 that satisfy the eligibility criteria of the program with the exception of that pertaining to ratings. Table 
\ref{tab:SummaryStatsIssuerCharacteristicsPreProgram} presents the summary statistics of the sample units.


\begin{table}[tbh]
	\caption{Summary statistics for the issuer characteristics.}
	\label{tab:SummaryStatsIssuerCharacteristicsPreProgram}
	\vspace{0.2em}
	\centering
	{\renewcommand{\arraystretch}{1.1}
		\begin{tabular}{@{}l*{7}{c}@{}}
			\hline\hline
			variable& definition																									& mean 	& sd 		& Q$_1$	& Q$_2$	& Q$_3$ & N  \\ \hline
			prof	& $\frac{\text{EBIT}}{\text{total revenue}}$										& 0.12	& 0.78	& 0.040	& 0.10	& 0.21	& 511\\		
			cf		& $\frac{\text{cash from operations}}{\text{total assets}}$			& 0.053	& 0.10	& 0.031	& 0.064	& 0.093	& 449\\
			liq		& $\frac{\text{cash from operations}}{\text{total liabilities}}$& 0.10	& 0.14	& 0.040	& 0.092	& 0.15	& 449\\
			cov		& $\frac{\text{EBIT}}{\text{interest expenses}}$								& 5.0		& 13		& 1.3		& 3.0		& 6.4		& 480\\
			lev		& $\frac{\text{total debt}}{\text{total assets}}$								& 0.39	& 0.21	& 0.23	& 0.37	& 0.54	& 489\\
			solv	& $\frac{\text{common equity}}{\text{total assets}}$						& 0.30	& 0.20	& 0.17	& 0.28	& 0.42	& 504\\
			size	& {\scriptsize $\log (\text{total revenue})$}										& 3.4		& 1.1		& 2.8		& 3.6		& 4.3		& 515\\
			age		& {\scriptsize 2017 \!$-$\! year founded}												& 71		& 71		& 20		& 54		& 101		& 453\\
			ltdebt	& $\frac{\text{long-term debt}}{\text{total assets}}$					& 0.34	& 0.28	& 0.15	& 0.28	& 0.45	& 491\\
			\hline
			\multicolumn{8}{@{}p{0.96\textwidth}@{}}{\footnotesize NOTE: The variable size is calculated with total revenue recorded in millions of euros.}
		\end{tabular}
	}
\end{table}

Figure \ref{fig:PS_preCSPP} illustrates the distribution of the propensity scores estimated from the ordered probit model for each rating category using the pre-CSPP sample.
\begin{figure}[tbh]
	\caption{Estimated propensity scores by rating in the pre-CSPP sample.}
	\begin{center}
	\includegraphics[scale=0.6,trim={0 1cm 0 2cm}]{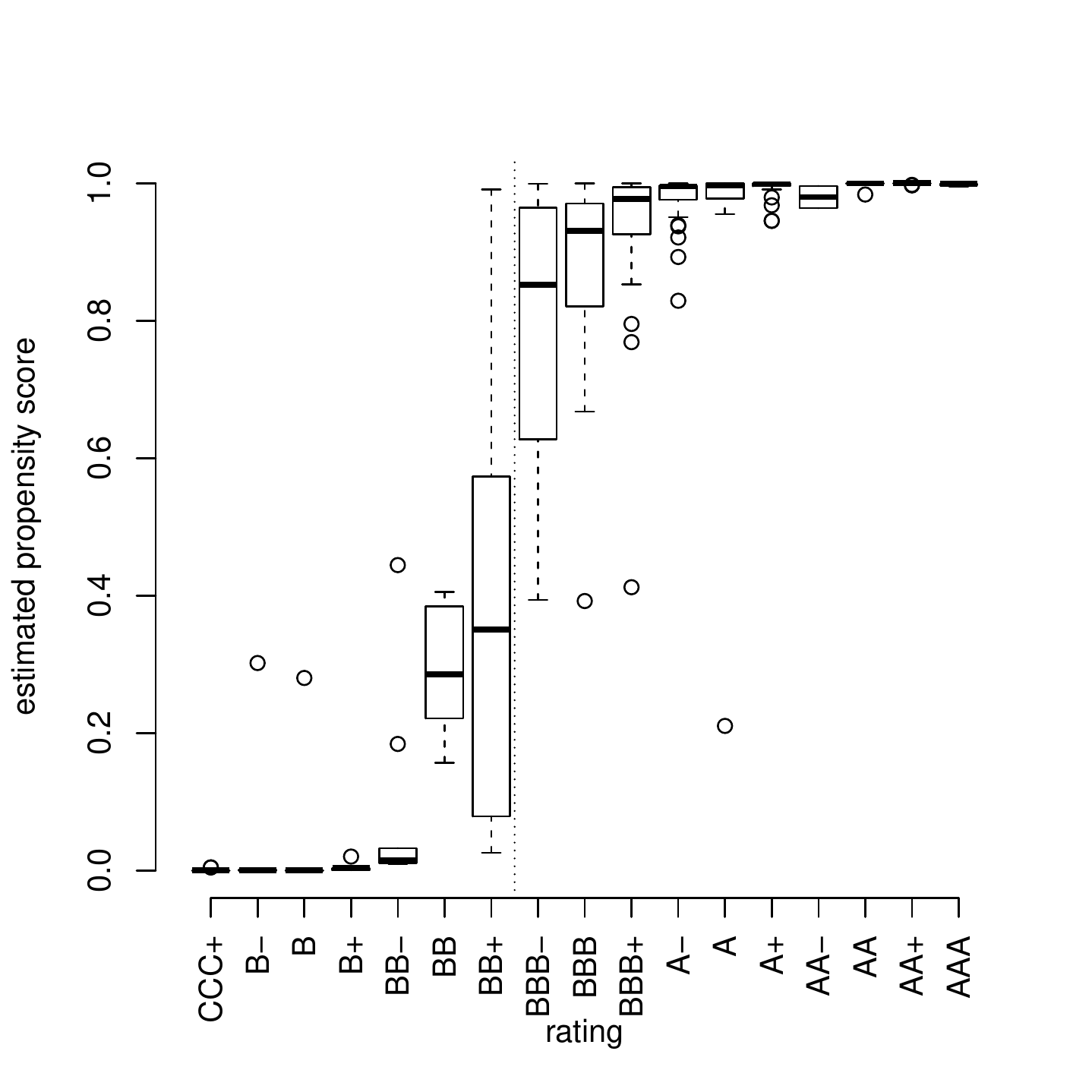}
	\end{center}
	\label{fig:PS_preCSPP}
\end{figure}

\bibliographystyle{jasa3}
\bibliography{references}

\end{document}